\documentclass[a4paper,11pt,aps,tightenlines,nofootinbib]{revtex4}
\usepackage{xspace}
\usepackage[english]{babel}
\usepackage{amsfonts}
\usepackage{amsmath, amsthm, amssymb, mathrsfs}
\usepackage[dvips]{graphicx}
\usepackage{pdfpages}
\usepackage{setspace}
\usepackage{indentfirst}
\usepackage[margin=2.5cm]{geometry}
\usepackage{bbm}

\usepackage{pstricks}
\usepackage{enumerate}
\usepackage{pstricks-add}
\usepackage{epsfig}
\usepackage{pst-grad}
\usepackage{pst-plot}

\newcommand{\abs}[1]{\left|#1\right|}
\newcommand{\bra}[1]{\left< #1 \right|}

\newcommand{\ket}[1]{\left|#1\right>}

\newcommand{\be}{\begin{equation}}
\newcommand{\ee}{\end{equation}}
\newcommand{\ba}{\begin{array}{c}}
\newcommand{\ea}{\end{array}}

\newcommand{\Hvtx}{$\mathcal{H}_{\text{vtx}}\,$}
\newcommand{\Cgr}{C^{\text{gr}}}
\def\beqr{\begin{eqnarray}}
\def\eeqr{\end{eqnarray}}

\newtheorem{req}{Requirement}

\numberwithin{equation}{section}
\numberwithin{thr}{section}
\numberwithin{chr}{section}
\numberwithin{df}{section}

\newcommand{\makeSymbol}[1]{\mathord{\vcenter{\hbox{#1}}}}

\begin{document}

\title{Hamiltonian operator for loop quantum gravity coupled to a scalar field}
\author{Emanuele Alesci}
\email[]{emanuele.alesci@fuw.edu.pl}
\author{Mehdi Assanioussi}
\email[]{mehdi.assanioussi@fuw.edu.pl}
\author{Jerzy Lewandowski}
\email[]{jerzy.lewandowski@fuw.edu.pl}
\author{Ilkka M\"{a}kinen,}
\email[]{ilkka.makinen@fuw.edu.pl}
\affiliation{Faculty of Physics, University of Warsaw, Pasteura 5, 02-093 Warsaw, Poland}

\begin{abstract}
We present the construction of a physical Hamiltonian operator in the deparametrized model of loop quantum gravity coupled to a free scalar field. This construction is based on the use of the recently introduced curvature operator, and on the idea of so-called {\it special loops}. We discuss in detail the regularization procedure and the assignment of the loops, along with the properties of the resulting operator. We compute the action of the squared Hamiltonian operator on spin network states, and close with some comments and outlooks.
\end{abstract}

\maketitle

\section{Introduction}
General relativity in Ashtekar-Barbero variables \cite{A variables, B variables} can be cast in an SU(2) Yang-Mills theory and treated as a Hamiltonian system with constraints consisting of the Gauss (gauge constraints), spatial diffeomorphism and Hamiltonian constraints. Canonical loop quantum gravity \cite{lqgcan1,lqgcan2,lqgcan3,lqgcan4}, which is an attempt of quantization a la Dirac \cite{DiracQM1964} of general relativity, has successfully completed the construction of a kinematical Hilbert space and the implementation of the Gauss constraints and the spatial diffeomorphism constraints \cite{Ashtekar:1995zh} in the quantum theory, leading to a gauge and spatial diffeomorphism invariant Hilbert space ${\cal H}_\text{Diff}^G$. The treatment of the last constraints is a more complicated task. The Hamiltonian has been regularized and promoted to an operator acting on ${\cal H}_\text{Diff}^G$ by Thiemann \cite{Thiemann96a} improving earlier attempts \cite{Early-hamiltonians}, however even if the general 
structure of the solutions to the Hamiltonian constraints is known, it is very difficult to define the physical Hilbert space. The issues are conceptual and technical.

Conceptual, because the Hamiltonian is not preserving ${\cal H}_\text{Diff}^G$ and even if attempts to deal with the absence of a physical Hilbert space have been explored \cite{Lewandowski:1997ba}, this problem has led to new research directions, in particular the master constraint program \cite{Thiemann:2003zv}, the algebraic quantum gravity program \cite{17}, the deparametrized models \cite{Brown-Kuchar, Rov-Smo, KucharRomano, Domagala:2010bm, Husain:2011tk, Giesel:2012rb} in the canonical setting, the spinfoam program \cite{lqgcov} in the covariant framework and also some toy models \cite{Laddha:2010wp, Henderson:2012ie, Henderson:2012cu, Laddha:2014xsa} in which an alternative quantization strategy of the Dirac algebra is applied. 

Concerning the technical difficulties, the Hamiltonian constraint is composed of two terms: the Euclidean part and the Lorentzian part. Both are non polynomial in the canonical variables, specially the second term that involves a double Poisson bracket of the Euclidean part with the volume and has a complicated form in terms of Ashtekar variables. A clever way to tame the non polynomial character of the constraints is using ``Thiemann's trick", i.e. replacing the classical non polynomial functions by Poisson brackets of polynomial functions with the volume and of the Euclidean part with the volume. Once promoted to an operator the resulting expression comprises several commutators containing the volume operator \cite{RovelliSmolin95,AshtekarLewand95, AshtekarLewand98}. While this procedure helps to bypass the non polynomial character of the constraint, the resulting operator however is not self-adjoint and the explicit calculation of the Hamiltonian action is impossible because the volume operator present in 
the final expression has no explicit spectral decomposition. The partially formal result is already an extremely involved expression \cite{Alesci:2011ia, Alesci:2013kpa}.

In this work we present another proposal for quantizing the Hamiltonian constraints. The first change is already in the classical formula for the scalar constraint. It is the sum of terms proportional to the Euclidean scalar constraint and, respectively, the Ricci scalar of the three metric tensor \cite{DL}. Our aim is to implement the dynamics in the quantum model of gravity coupled to a free scalar field \cite{Domagala:2010bm}. The construction is conceptually based on the recently introduced ``intermediate'' Hilbert space \Hvtx \cite{Lewandowski:2014hza} that is preserved by the obtained Hamiltonian operator, raising hope for a well defined evolution operator with satisfactory properties, e.g. self adjointness.

The developed regularization is based on a concrete implementation of a proposal first appeared in \cite{RovSmo} concerning the Euclidean constraint, and the use of the curvature operator introduced in \cite{Curvature_op.} to deal with the Lorentzian part. The paper is organized as follows. In section \ref{section_1} we review the classical model of gravity minimally coupled to a scalar field; in section \ref{section_2} we review the loop quantum gravity construction, present the regularization of the Hamiltonian and discuss the quantum operator and its properties; then we close in section \ref{section_3} with some conclusions and outlooks to further developments of this program.

\section{Classical theory}\label{section_1}

Considering gravity minimally coupled to a scalar field in the standard ADM formalism \cite{ADM}, the theory is set as a constrained system for the standard canonical variables $q_{ab}(x)$ and $\phi(x)$, respectively the metric and the scalar field on a $3d$ manifold $\Sigma$ with conjugate momenta $p^{ab}(x)$ and $\pi(x)$. The analysis shows that the vector constraints $C_a(x)$ and the scalar constraints $C(x)$ in this model are expressed in terms of the vacuum gravity constraints, $\Cgr_a(x)$ and $ \Cgr(x)$, and the scalar field variables as follows
\be
C_a(x)\ =\  \Cgr_a(x) \ +\ \pi(x)\phi_{,a}(x)\label{vector},
\ee
\be
C(x)\ =\Cgr(x)\ +\ \frac{1}{2}\frac{\pi^2(x)}{\sqrt{q(x)}} +\frac{1}{2}q^{ab}(x)\phi_{,a}(x)\phi_{,b}(x)\sqrt{q(x)} + V(\phi)\sqrt{q(x)}\label{scalar},
\ee
where $q$ is the determinant of the metric $q_{ab}$.\\

With the Ashtekar-Barbero variables $(A^i_a, E_i^a)$ ($i=1,2,3$) used in LQG, 
\begin{align}
\{A_a^i(x),E^b_j(y)\}\ =&\ 8\pi\beta G\delta_a^b\delta^i_j\delta(x,y)\\
\{A_a^i(x),A_b^j(y)\}\ =\ &0\ =\ \{E^a_i(x),E^b_j(y)\}
\end{align}
where $G$ is Newton constant and $\beta$ is the Immirzi parameter, additional constraints -- the Gauss constraints generating Yang-Mills gauge transformations -- are induced:
\be
G^i(x)\ = \partial_a E_i^a + \epsilon_{ij}{}^k A^j_a E^a_k.\label{gauss}
\ee
The field $A^i_a$ is identified with an su(2)- valued differential 1-form
\be
A\ =\ A^i_a\tau_i\otimes dx^a
\ee
while the field $E^a_i$ with an su(2)${}^*$ vector density
\be
E\ =\ E_i^a\tau^*{}^i\otimes \frac{\partial}{\partial x^a} 
\ee
where $\tau_1,\tau_2,\tau_3\in$su(2) is a basis of su(2) such that
\be
-2 \,{\rm Tr}\,\tau_i\tau_j\ =\ \delta_{ij}.
\ee
A solution by points in the phase space $(A^i_a,E_i^a,\phi,\pi)$ must satisfy all the constraints:
\be 
G^i(x)=0 \qquad \qquad C_a(x)=0 \qquad \qquad C(x)=0  \label{soluzioni}.
\ee
In terms of the Ashtekar-Barbero variables, the gravitational part of the scalar constraint reads
\be\label{Cgr} 
C^{\rm gr}(x) = -\frac{1}{16\pi\beta^2G} \biggl(\frac{\epsilon_{ijk}E^a_i(x)E^b_j(x)F_{ab}^k(x)}{\sqrt{|\det E(x)|}} + (1+\beta^2)\sqrt{|\det E(x)|} \,R(x)\biggr)
\ee
where $R$ is the Ricci scalar of the metric tensor $q_{ab}$ on $\Sigma$ related to the Ashtekar frame variable by
\be  q^{ab}\ =\ \frac{E^a_iE^b_i}{{|\det E|}} .\ee   
The first term of  $C^{\rm gr}$ is usually related to the Euclidean scalar constraint
\be\label{Eucclass}
C^{\rm Eucl}\ :=\ -\frac{1}{16\pi G \beta^2}  \frac{\epsilon_{ijk}E^a_i(x)E^b_j(x)F_{ab}^k(x)}{\sqrt{|\det E(x)|}}  .
\ee

To construct a quantum theory mainly two strategies can be adopted. The first is to promote the whole set of constraints to operators defined in an appropriate Hilbert space and look for the states annihilated by the constraints operators to build a Physical Hilbert space. The second, that we consider in this work, is to deparametrize the theory classically then quantize. The deparametrization procedure starts with assuming that the constraints \eqref{soluzioni} are satisfied, hence we can solve the vector constraints for the gradient of the scalar field,
\be
\phi_{,a}=-\frac{\Cgr_a}{\pi},
\ee
and then use this condition in \eqref{scalar} to solve it for $\pi$:
\be
\pi^2\ =\ \sqrt{q}\left( -\left(\Cgr+\sqrt{q}V(\phi)\right)\pm \sqrt{\left(\Cgr+\sqrt{q}V(\phi)\right)^2-q^{ab}\Cgr_a\Cgr_b} \right).\label{pm}
\ee
In case of vanishing potential 
\be
V(\phi)=0,
\ee 
which is our assumption in the rest of this article, equation \eqref{pm} represents the deparametrization of the system with respect to the scalar field, which can be seen as an emergent time. Note that in this case, on the constraint surface, it is necessary to have
\be 
\Cgr(x)\ \leq\ 0. 
\ee
The sign ambiguity in (\ref{pm}) amounts to treating different regions of the phase space, namely for $+$ and $-$ respectively
\be
\pi^2\ \ge / \le \ q^{ab}(x)\phi_{,a}(x)\phi_{,b}(x)\,q(x).
\ee

We choose the phase space region corresponding to $+$ and $\ge$. It contains spacially homogeneous spacetimes useful in
cosmology. Then, the scalar constraints can be rewritten in an equivalent form as
\beqr
C'(x)\ &=&\ \pi(x)\ \mp\ h(x),\label{newscalar}\\
h\ &:=&\ \sqrt{-\sqrt{q}C^{gr}+\sqrt{q}\sqrt{({C^{gr}})^2-q^{ab}C^{gr}_aC^{gr}_b}}.
\label{hclassic}
\eeqr
We will also restrict ourselves to the case of
\be 
\pi(x)\ \ge\ 0, 
\ee
although technically there is no problem in admitting both signs in the quantum theory.\\
 
The constraints $C'$ commute strongly,
\begin{align}
\{C'(x),C'(y)\}=0,
\end{align}
implying \cite{KucharRomano}
\begin{align}
\{h(x),h(y)\}=0.
\end{align}

In this case a Dirac observable $\mathscr{O}$ on the phase space would satisfy
\begin{align}
\{\mathscr{O},G^i(x)\}=\{\mathscr{O},C_a(x)\}=\{\mathscr{O},C'(x)\}=0.
\end{align}
The vanishing of the first and second Poisson brackets induce gauge invariance and spatial diffeomorphism invariance respectively. The vanishing of the third Poisson bracket is equivalent to writing
\begin{align}
 \frac{\partial \mathscr{O}}{\partial \phi}\ =\ \{\mathscr{O},\pi(x)\}\ =\ \{\mathscr{O},h(x)\},
\end{align}

\section{Quantum theory}\label{section_2}
\subsection{The general structure}
The quantization of gravity coupled to a massless scalar field was performed in \cite{Domagala:2010bm,Domagala:2011}. While the derivation was partially formal - the existence of the operators $\hat{C}^{\rm gr}_a$ is assumed at some stage - the result is expressed in a derivable way by elements of the framework of LQG:
\begin{itemize}
\item The physical Hilbert space ${\cal H}$ is the space of the  quantum states of the {\it matter free}     
gravity which satisfy the quantum vector constraint and the quantum Gauss constraint.    

\item The dynamics  is defined by a Schr\"{o}dinger like equation
\be \frac{d}{d t} \Psi\ =\ -\frac{i}{\hbar}\hat{H}\Psi\ 
\ee
where $t$ is a parameter of the transformations 
$$\varphi\mapsto \varphi+t\ .$$
\item The quantum Hamiltonian 
\be\label{H}
\hat{H}\ =\ \int d^3x \widehat{\sqrt{-2\sqrt{q(x)}C^{\rm gr}(x)}}  
\ee  
is a quantum operator corresponding to the classical  observable
\be
{H}\ =\ \int d^3x {\sqrt{-2\sqrt{q(x)}C^{\rm gr}(x)}}. 
\ee
\end{itemize}
This operator could be defined by using already known operators $\widehat{\sqrt{q(x)}}$ and $\widehat{C^{\rm gr}(x)}$, as outlined in \cite{Domagala:2011}. However, the observable $\sqrt{q}C^{\rm gr}$  written in terms of the Ashtekar-Barbero variables reads
\be\label{h(x)}
-\sqrt{|{\rm det}E(x)|}C^{\rm gr}(x) = \frac{1}{16\pi\beta^2G} \biggl(\epsilon_{ijk}E^a_i(x)E^b_j(x)F_{ab}^k(x) + (1+\beta^2){|{\rm det} E}(x)| \,R(x).\biggr)
\ee
The denominator ${\sqrt{|\det E(x)|}}$ present in (\ref{Cgr}) disappears in (\ref{h(x)}). Moreover, the formula (\ref{reg_L}) below for $\sqrt{{|{\rm det} E}(x)}| \,R(x)$ expressed in terms of the quantizable observables (holonomies and fluxes) also contains the same denominator, which again disappears after using the formula (\ref{h(x)}). That coincidence of reductions motivates us to quantize  the expression (\ref{h(x)}) for $h(x)$ directly.

\subsection{Kinematical Hilbert space}
The  kinematical quantum states in LQG are \emph{cylindrical} functions of the variable $A$, i.e., they depend on $A$ 
only through finitely many parallel transports 
\begin{equation} h_e[A]\ =\ {\rm P}\exp\left(-\int_e A \right) \end{equation}
where $e$ ranges over finite curves -- we will also refer to them as  \emph{edges} -- in $\Sigma$. 
That is a kinematical quantum state $\Psi$ has the form
\begin{equation}\label{cyl'} 
\Psi[A]\ =\ \psi(h_{e_1}[A],\ldots ,h_{e_n}[A]) 
\end{equation} 
with a function $\psi\ :\ {\rm SU}(2)^n \ \rightarrow \mathbb{C}$. Here, for every edge we choose an orientation 
to define the parallel transport $h_{e_I}[A]$. 

To calculate the scalar product between two cylindrical functions $\Psi$ and $\Psi'$ we find in $\Sigma$ an embedded graph $\gamma''=\{e''_1,\ldots ,e''_{n''}\}$, such that both functions can be written as \footnote{The existence of $\gamma''$
is ensured by assuming the analyticity of $\Sigma$ and of the edges of the graphs \cite{lqgcan1}.}
\begin{equation}
\begin{split}
\Psi[A] &= \psi(h_{e''_1}[A],\ldots ,h_{e''_{n''}}[A]),\\
\Psi'[A] &= \psi'(h_{e''_1}[A],\ldots ,h_{e''_{n''}}[A]).
\end{split}
\end{equation}
The scalar product is
\begin{equation}\label{(|)} (\Psi | \Psi')\ =\ \int dg_1\ldots dg_{n''}\overline{\psi(g_1,\ldots ,g_{n''})}{\psi}'(g_1,\ldots ,g_{n''}).\end{equation}
We denote the space of all the cylindrical functions defined as above with a graph $\gamma$ by ${{\rm Cyl}}_\gamma$ and, 
respectively, the space of all cylindrical functions by ${\rm Cyl}$. The kinematical Hilbert space ${\cal H}_{\rm kin}$ is the completion 
\begin{equation} {\cal H}_{\rm kin} \ =\ \overline{{\rm Cyl}}\end{equation}
with respect to the Hilbert norm defined by (\ref{(|)}).

Every cylindrical function $f$ is also a quantum operator
\begin{equation}\label{hol-op}(\widehat{f(A)}\Psi')[A]\ =\ f[A]\Psi'[A].\end{equation}
A typical  example is  
\be
f(A) = D^{(j)}{}^a_{\ b}(h_p(A)) 
\ee    
defined by a path $p$ in $\Sigma$, a half-integer $j=0,\frac{1}{2},1,\frac{3}{2},...$, the corresponding representation 
\be D^{(j)}:{\rm SU(2)}\rightarrow {\rm Unitary}({\cal H}^{(j)})  \ee
and some orthonormal basis $v_1,...,v_{2j+1}\in {\cal H}^{(j)}$, 
\be
D^{(j)}{}^a_{\ b}(g)\ :=\ (v_a\,|\,D^{(j)}(g)v_b)_{{\cal H}^{(j)}} .
\ee
 
Note that a connection operator ``$\widehat{A}$'' itself is not defined.\\

An operator $\hat{J}_{x[e]\xi}$, which is naturally defined in this framework, is assigned to a triple $(x,\xi, [e])$, where  $x\in\Sigma$, $\xi\in$ su(2) and $[e]$ is a maximal family of curves beginning at $x$ such that each two curves overlap on a connected initial segment containing $x$. To define the action of $\hat{J}_{x[e]\xi}$ on a function $\Psi\in {\rm Cyl}$, we represent this function on a graph such that $e_I\in [e]$. The action is
\be
\hat{J}_{x[e]\xi}\Psi\ =\ i\hbar\frac{d}{d\epsilon}\bigg|_{\epsilon=0}\psi(h_{e_1}e^{\epsilon\xi},h_{e_2},...,h_{e_n}).
\ee

For $\xi=\tau_i$, it is convenient to introduce a simpler notation  
\be
\hat{J}_{x,e,i}\ :=\ \hat{J}_{x[e]\tau_i}.
\ee

The field $E^a_i(x)$ is naturally quantized as  
\begin{align}\widehat{E}^a_i(x)\Psi[A]\ =\ \frac{\hbar}{i}\{\Psi[A],E^a_i(x)\} \ =\ \frac{8\pi\beta l_\text{P}^2}{i}
\frac{\delta}{\delta A^i_a(x)}\Psi[A]. 
\end{align}

Given an edge $e:[t_0,t_1]\rightarrow\Sigma$, and a function $f\in C(SU(2))$, the variation is given by the following formula
\be
\frac{\delta}{\delta A^i_a(x)} f(h_e(A))\ =\ -\int_{t_0}^{t_1}dt\dot{e}^a(t) \delta(x,e(t))\left(h_{e,t_1,t}(A)\tau^ih_{e,t,t_0}(A)\right)^A_B
\frac{\partial}{\partial g^A_B}f(g)_{|_{g=h_e(A)}}  ,    
\ee 
where by $h_{e,t_1,t}(A)$ (respectively, $h_{e,t,t_0}(A)$) we mean the parallel transport with respect to $A$ along $e$ from the point 
$e(t)$ to $e(t_1)$ ($e(t_0)$ to $e(t)$),
and by the partial derivatives with respect to group elements we mean
\be
\frac{d}{d\epsilon}{\bigg|_{\epsilon=0}}f(ge^{\epsilon \xi})\ :=\ (g\xi)^A_B\frac{\partial}{\partial g^A_B}f(g) . 
\ee 
Smearing along 2-surfaces  leads to  well defined operators in ${\cal H}_{\rm kin}$. Let $S\subset \Sigma$ be an oriented, 
2-dimensional surface, and 
\be \xi:S\rightarrow {\rm su(2)}  
\ee
be a smearing function. The flux corresponding to $E$ is
\begin{align}\label{flux-clas} {P}_{S,\xi}\ :=\ \int_S \frac{1}{2}dx^b\wedge dx^c\epsilon_{abc}\xi^i(x) {E}^a_i(x) .
\end{align}
The quantum flux is a well defined operator
\begin{align}\label{flux-op} \hat{P}_{S,\xi}\  =\ 
4\pi G\sum_{x\in S}\xi^i(x)\sum_{e}\kappa_S(e)\hat{J}_{x,e,i}
\end{align}
where $e$ runs through the classes of curves beginning at $x$, and
\be
\kappa_S(e)\ =\ -1,0,1, 
\ee 
depending on whether $e$ goes down, along, or, respectively, up the surface $S$. A generalized function  $\xi$ may also involve parallel transports depending on $A$. A typical 
example is
\be\label{partrFl}
\xi(x)\ =\ {\rm Ad}(h_{p(x)}(A))\zeta, \ \ \ \ \ \ \zeta\ \in\ {\rm su(2)}   
\ee
where
\be x\mapsto p(x) \ee
assigns  to each point $x$ a path $p(x)$,  $h_{p(x)}(A)$ is the parallel transport, and  Ad is the adjoint action of SU(2) in the Lie algebra su(2)
\be
{\rm Ad}(g)\zeta\ =\ g\zeta g^{-1} .
\ee

In conclusion, the operators compatible with the LQG structure of ${\cal H}_{\rm kin}$ are (functions of the) parallel transports and fluxes.\\

The quantum Gauss constraint operator reads
\be  \hat{G}_i(x)\ =\  \sum_{y\in\Sigma} \delta(x,y)\sum_{e\ \text{at}\ y}\hat{J}_{y,e,i}.  \ee
Solutions in ${\rm Cyl}$ to the Gauss constraint
\be  \hat{G}_i(x)\Psi\ =\ 0  \ee
are functions such that 
\be f(A)\ =\ f(g^{-1}Ag + g^{-1}dg),\ \ \  {\rm for \ every}\ \ g \in C^1(\Sigma,{\rm SU(2)}) .\ee
We denote their algebra, subalgebra of Cyl by Cyl$^G$, and the corresponding subspace of ${\cal H}_{\rm kin}$ by    ${\cal H}^G_{\rm kin}$.  
A dense subspace  of  ${\cal H}^G_{\rm kin}$  is spanned by the spin network functions. A spin network function is defined by a graph $\gamma=(e_1,...,e_n)$, half integers (non zero) $(j_1,...,j_n)$ assigned to the edges and  intertwiners $(\iota_1,...,\iota_m)$ assigned to the vertices $(v_1,...,v_m)$:
\be  \Psi(A)\ =\  D^{(j_1)}{}^{a_1}_{\ b_1}(h_{e_1}(A))\ ...\   D^{(j_n)}{}^{a_n}_{\ b_n}(h_{e_n}(A))
( \iota_1\otimes\ ...\ \otimes\iota_m)_{a_1...a_n}^{b_1...b_n} . \ee 
Each $\iota_\alpha$ is an invariant of the tensor product of the representations assigned to the edges $e_I$ whose source is $v_\alpha$ and the representations dual to those assigned to the edges whose target is $v_\alpha$.\\

Given a graph $\gamma$, we denote by $\rm Cyl_\gamma^G$ the space spanned by all the spin network functions 
defined on this graph, and 
\be  {\cal H}^G_\gamma\ :=\ \overline{\rm Cyl_\gamma^G} .\ee     
  
To define the orthogonal decomposition of the space of the Gauss constraint solutions we need to admit closed edges, that is edges for which the end point equals the beginning point, and closed edges without vertices (embeddings of a circle in $\Sigma$). In the case of an edge without vertices, we choose a beginning-end point arbitrarily in the definition of the spin network function. On the other hand we do not count those graphs that can be obtained from another graph by the splitting of an edge. Then the space of all the solutions to the Gauss constraint can be written as the orthogonal sum
\be\label{decomp} {\cal H}_{\rm kin}^G\ =\ \overline{\bigoplus_{\gamma} {\cal H}^G_\gamma}\ee
where $\gamma$ ranges over all the un-oriented graphs defined in this paragraph.

\subsection{The vertex Hilbert space}
\label{se_new}
Every analytic diffeomorphism $f\in $ Diff$^\omega$($\Sigma$) defines a unitary operator 
$U_{f}:{\cal H}_{\rm kin}\rightarrow {\cal H}_{\rm kin}$,
\begin{equation}\label{UfPsi} 
U_f\Psi[A]\ =\ \Psi[f^*A]. 
\end{equation}
Given a graph $\gamma$  consisting of edges and vertices
\begin{equation*}
{\rm Edge}(\gamma):=\{e_1,\ldots ,e_n\},\quad
{\rm Vert}(\gamma)=\{v_1,\ldots ,v_m\}, 
\end{equation*}
the action of $U_f$ on a cylindrical function (\ref{cyl'}) reads
\begin{equation}\label{UfPsi'} 
U_f\Psi[A]\ =\ \psi(h_{f(e_1)}[A],\ldots ,h_{f(e_n)}[A]),
\end{equation}
where for the parallel transport along each edge $f(e_I)$ we choose the orientation induced by the map $f$ and the orientation of $e_I$ chosen in (\ref{cyl'}). Smooth diffeomorphisms map analytic graphs into smooth graphs, therefore their action is not defined in our Hilbert space ${\cal H}_{\rm kin}$. Suppose, however,  that given a graph $\gamma$, a smooth diffeomorphism $f\in $ Diff$^\infty(\Sigma$) maps $\gamma$  into an analytic graph. Then (\ref{UfPsi}) and (\ref{UfPsi'}) define a unitary map
\begin{equation}\label{Diffinfty}
U_f\ : {\cal H}_\gamma\ \rightarrow\   {\cal H}_{f(\gamma)} .
\end{equation}

The idea of the vertex Hilbert space of \cite{Lewandowski:2014hza} is to construct from  elements of the Hilbert space
${\cal H}^G_{\rm kin}$  partial solutions to the vector constraints, by averaging the elements of each of the sub-spaces ${\cal H}^G_{\gamma}$ with respect to  all the smooth diffeomorphisms Diff$^\infty(\Sigma)_{{\rm Vert}(\gamma)}$ which act trivially in the set of the vertices ${\rm Vert}(\gamma)$. Denote by TDiff$^\infty(\Sigma$)$_\gamma$ the subset of Diff$^\infty(\Sigma$) which consists of all the diffeomorphisms $f$ such that $f(\gamma)=\gamma$ and $U_f$ acts trivially in  ${\cal H}^G_{\gamma}$. 

Denote by Diff$^\infty_\gamma(\Sigma)_{{\rm Vert}(\gamma)}$ the set of those elements of Diff$^\infty(\Sigma)_{{\rm Vert}(\gamma)}$ which preserve the analyticity of $\gamma$. The maps 
${\cal H}^G_{\gamma}\longrightarrow {\cal H}^G_{\rm kin}$
obtained by the diffeomorphisms in Diff$^\infty_\gamma(\Sigma$)$_{{\rm Vert}(\gamma)}$ are in one to one correspondence with  the elements of the quotient 
\begin{equation} 
{\rm D}_\gamma := {\rm Diff}^\infty_\gamma(\Sigma)_{{\rm Vert}(\gamma)} / {\rm TDiff}(\Sigma)_\gamma. 
\end{equation}
Since ${\rm D}_\gamma$ is a non-compact set and we do not know any probability measure on it, we define the averaging in  ${\rm Cyl}^*$, the algebraic  dual to ${\rm Cyl}$. Given $\Psi \in {\cal H}^G_\gamma$, we turn it into  $\langle\Psi| \in {\rm Cyl}^*$,
$$ \bra\Psi   \ :\ \Psi'\in{\rm Cyl}^G:=\bigoplus_{\gamma'}{\rm Cyl}_{\gamma'}^G \ \mapsto (\Psi | \Psi'),$$
and average in ${\rm Cyl}^*$,
\begin{equation}\label{eta}
 \eta(\Psi)\ =\ \frac{1}{N_\gamma} \sum_{[f]\in D_\gamma}
\bra{U_f\Psi} , 
\end{equation} 
where
\begin{align}N_\gamma\ &= \  |{\rm Sym}_\gamma|\nonumber\\
{\rm Sym}_\gamma\ &:=\ \{f\in  {\rm Diff}^\infty(\Sigma)_{\{x_1,\ldots ,x_m\}}\ :\ f(\gamma)=\gamma\}\, /\, {\rm TDiff}(\Sigma)_\gamma .
\end{align} 
The resulting $\eta(\Psi)$ is a well defined linear functional 
\begin{equation*}
\eta(\Psi): {\rm Cyl}^G \rightarrow \mathbb{C}
\end{equation*}
because given $\Psi'\in{\rm Cyl}^G$, only a finite set of terms in the sum contribute to the number $\eta(\Psi)(\Psi')$. Hence we have defined a map
\begin{equation*} {\cal H}^G_\gamma \ni \Psi\ \mapsto\  \eta(\Psi)\ \in\ {\rm Cyl}^*\end{equation*} 
for every embedded graph $\gamma$.  We extend it by linearity to the algebraic orthogonal sum 
(\ref{decomp})
\begin{equation}
\eta:{\cal H}_\text{kin}^G \longrightarrow {\rm Cyl}^*.
\end{equation}

The vertex Hilbert space  ${\cal H}^G_{\rm vtx}$ is defined as the completion
\begin{equation}{\cal H}^G_{\rm vtx}\ :=\ \overline{\eta({\rm Cyl}\cap {\cal H}^G_{\rm kin})}  
\end{equation}
under the norm induced by the natural scalar product
\begin{equation} \label{scpr}
(\eta(\Psi)|\eta(\Psi'))\ :=\ \eta(\Psi)(\Psi'). 
\end{equation} 
It has an orthogonal decomposition that is reminiscent of 
\eqref{decomp}:
 Let ${\rm FS}(\Sigma)$ be the set of finite subsets of $\Sigma$. Then
\begin{align} \label{decomp1}
{\cal H}^G_{\rm vtx}\ &=\ \overline{\bigoplus_{V\in {\rm FS}(\Sigma)}{\cal H}^G_{V}}\\
\label{decomp2}
{\cal H}^G_{V}\ &:=\ \overline{\bigoplus_{[\gamma]\in[\gamma(V)]}\mathcal{H}^G_{[\gamma]}}\\
{\cal S}^G_{[\gamma]}\ &:= \ \eta({\cal S}^G_\gamma)
\end{align}
where  $\gamma(V)$ is the set of graphs $\gamma$ with vertex set $V={\rm Vert}(\gamma)$, 
$[\gamma(V)]$ is the set of the 
Diff$(\Sigma)_V$-equivalence classes $[\gamma]$ of the graphs $\gamma\in \gamma(V)$ and ${\cal S}^G_\gamma$  is the subspace ${\cal S}^G_\gamma\subset {\cal H}^G_\gamma$ of  the elements  invariant with respect to the symmetry group ${\rm Sym}_\gamma$. Importantly, 
\be
\eta: {\cal S}^G_\gamma\ \rightarrow \ {\cal S}^G_{[\gamma]}
\ee
is an isometry. The orthogonal complement of ${\cal S}^G_\gamma$ in ${\cal H}^G_\gamma$, on the other hand, 
is annihilated by $\eta$.\\      
 
The Hilbert space ${\cal H}^G_{\rm vtx}$ carries a natural action of Diff$^\omega(\Sigma)$, which we will also denote by $U$. 
It is defined by
\begin{equation}
\label{diff}
U_f \eta(\Psi):=\eta(U_{f} \Psi), \qquad f\in {\rm Diff}^\omega(\Sigma)
\end{equation}
A short calculation shows that $U_f$ is unitary and maps $\mathcal{H}^G_V$ to $\mathcal{H}^G_{f(V)}$ in the decomposition \eqref{decomp1}.

Each subspace ${\cal S}^G_{[\gamma]}$ consists of Diff$^\omega(\Sigma)_{{\rm Vert}(\gamma)}$ invariant elements. In this sense, they are partial solutions to the quantum vector constraint. They can be turned into full solutions of the quantum vector constraint by a similar averaging with respect to the remaining Diff($\Sigma$)/Diff($\Sigma$)$_{{\rm Vert}(\gamma)}$ \cite{Lewandowski:2014hza}. We denote the space of those solutions ${\cal H}_\text{Diff}^G$.
 
\subsection{The Hamiltonian operator}
In the Hilbert space ${\cal H}^G_{\rm vtx}$ we will introduce (derive) an operator
\be\label{hat-H}
\hat{H}\ =\ \widehat{\int_\Sigma d^3x\,\sqrt{-2\sqrt{q(x)}C^{\rm gr}(x)}}, 
\ee
where we have 
\be
-2\sqrt{|{\rm det}E(x)|}C^{\rm gr}(x) = 
\frac{1}{8\pi\beta^2G} \biggl(\epsilon_{ijk}E^a_i(x)E^b_j(x)F_{ab}^k(x) + (1+\beta^2){|{\rm det} E}(x)| \,{}^{(3)}R(x).\biggr)
\ee
In order to define the corresponding operator, we need to consider how to regularize and quantize an expression of the form
\be\label{A2+B2}
\int_\Sigma d^3x\,f(x)\sqrt{a^2(x) + b^2(x)}
\ee
where $f$ is a smearing function defined on $\Sigma$ while $a(x)$ and $b(x)$ are functionals of the  fields
$A^i_a$ and $E^a_i$.
  
Introducing a decomposition of the manifold $\Sigma$ into cells $\Delta$, the integral can be approximated as
\be\label{reg A2+B2}
\int_\Sigma d^3x\,f(x)\sqrt{a^2(x) + b^2(x)} = \sum_\Delta \sqrt{\biggl(\int_\Delta d^3x\,f(x)a(x)\biggr)^2 +  
\biggl(\int_\Delta d^3x\,f(x)b(x)\biggr)^2} + O(\epsilon_\Delta),
\ee
where for every cell $\Delta$, $\epsilon_\Delta^3$ denotes the coordinate volume of  $\Delta$. 
If the integrals $\int_\Delta d^3x\,f(x)a(x)$ and $\int_\Delta d^3x\,f(x)b(x)$ can be quantized as well-defined operators, equation \eqref{reg A2+B2} then shows how to define the operator corresponding to $\int_\Sigma d^3x\,f(x)\sqrt{a^2(x) + b^2(x)}$. Equation \eqref{reg A2+B2} is the basis of our construction of the operator \eqref{hat-H}.

In our case, the operators corresponding to $a(x)$ and $b(x)$ themselves will be available, and will have the general form
\be\label{Ax and Bx}
\hat a(x) = \sum_{v\in\Sigma} \delta(x,v)\hat a_v, \qquad \hat b(x) = \sum_{v\in\Sigma} \delta(x,v)\hat b_v,
\ee
where the operators $\hat a_v$ and $\hat b_v$, when applied to a spin network state defined on a graph, have a non-zero action 
only if $v$ is one of the vertices of the graph. In this case, the operator
\be
\widehat{\int_\Sigma d^3x\,f(x)\sqrt{a^2(x) + b^2(x)}}
\ee
can be defined simply by inserting $\hat a(x)$ and $\hat b(x)$ into the right-hand side of equation \eqref{reg A2+B2}. In this way one obtains an operator, whose restriction to the space of spin network states defined on a given graph $\gamma$ takes the form 
\be
\widehat{\int_\Sigma d^3x\,f(x)\sqrt{a^2(x) + b^2(x)}} \ \bigg|_{{\cal H}^G_\gamma} = 
\sum_{v\in\gamma} f(v)\sqrt{\hat a_v^2 + \hat b_v^2}.
\ee
In other words, our regularization gives
\be
\widehat{\sqrt{a^2(x) + b^2(x)}} \ =\ \sum_{v\in\Sigma}\delta(x,v)\sqrt{\hat{a}_v^2+\hat{b}_v^2}. 
\ee

\subsubsection{Euclidean part}\label{EuclideanPart}
We start with the quantization of the  Euclidean  part  of our Hamiltonian 
(see (\ref{Eucclass})). In equation \eqref{reg A2+B2}, the role of $a(x)$ is now played by the function
\be
H^{\rm Eucl}(x)\ :=\ \frac{1}{\sqrt{8\pi G \beta^2}}{\sqrt{\epsilon_{ijk}E^a_i(x)E^b_j(x)F_{ab}^k(x)}}.
\ee

Consequently, we consider the quantization of the integral
\be\label{CF}
\int d^3 x\,f(x) H^{\rm Eucl}(x)
\ee 
(where an arbitrary smearing function $f$ has been introduced).

According to the general framework of LQG, we need to express the integral in terms of parallel transports $h_e$ and fluxes $P_{S,i}$. The easiest example is to consider the Riemann sum for this integral obtained by considering a cubic partition ${\cal P}_\epsilon$ of $\Sigma$ into cells $\Box$ of coordinate volume $\epsilon^3$
\be
\epsilon^3\sum_{\Box}  f(x_\Box){\sqrt{\epsilon_{ijk}E^a_i(x_\Box)E^b_j(x_\Box)F_{ab}^k(x_\Box)}},  
\ee
and to distribute the $\epsilon$ suitably
\be
\sum_{\Box}  f(x_\Box){\sqrt{\epsilon_{ijk}(\epsilon^2E^a_i(x_\Box))(\epsilon^2E^b_j(x_\Box))(\epsilon^2F_{ab}^k(x_\Box))}}. 
\ee
For each cube $\Box$ denote by $x_\Box$ the center, by $S^a_\Box$, $a=1,2,3$, three sides $x^a=$const 
(for each $a$ there are two, choose any one and orient such that the following is true). Moreover, 
for every $x\in \Box$, denote by  $p_\Box(x)$ the line from $x_\Box$ to $x\in \Box$.  Then, we have
\begin{align}
\epsilon^2 E^a_i\ &=\ P_{S^a_\Box,i} + o(\epsilon^2)\nonumber\\
\epsilon^2 F^k_{ab}\ &=\ \epsilon_{abc}(h_{\partial S^c_\Box}^k)^{(l)} + o(\epsilon^2),  
\end{align} 
where by $P_{S^a_\Box,i}$ we mean $P_{S,\xi}$ of (\ref{flux-clas}) with 
\be
S=S^a_\Box, \ \ \ {\rm and}\ \ \  \xi(x)\ :=\ h_{p_\Box(x)}\tau_ih_{p_\Box(x)}^{-1}\ ,
\ee
$h_{p_\Box(x)}$ standing for the parallel transport (with respect to a given field $A$) along $p_\Box$, 
and for a $SU(2)$ element $h$ we define\footnote{Equation \eqref{h^k} is obtained using the relations 
\[ D^{(l)}(h) = h^0\mathbbm{1}^{(l)} + h^k(\tau_k)^{(l)}, \qquad {\rm Tr}\,(\tau_i)^{(l)} = 0, \qquad {\rm Tr}\,\bigl((\tau_i)^{(l)}(\tau^k)^{(l)}\bigr) = \frac{W_l^2}{3}\delta_i^k. \]}
\be\label{h^k}
(h^k)^{(l)} = \frac{3}{W_l^2}\text{Tr}\,\Bigl(D^{(l)}(h)(\tau^k)^{(l)}\Bigr)
\ee
with $W_l=i\sqrt{l(l+1)(2l+1)}$. In this way we write the original expression in terms of fluxes and parallel transports (as a limit),
\be
\int d^3 x f H^{\rm Eucl {\cal P}_\epsilon}\ =\ \frac{1}{\sqrt{8\pi G \beta^2}}\sum_{\Box}  f(x_\Box)\sqrt{\epsilon_{ijk}P_{{S^a_\Box} ,i} P_{{S^b_\Box},j} \epsilon_{abc}(h_{\partial S^c_\Box }^k)^{(l)}},
\ee  
in the sense that
\be
\int d^3 x f H^{\rm Eucl {\cal P}_\epsilon}\ \rightarrow\  \int d^3 x f H^{\rm Eucl},
\ee
in the limit $\epsilon\rightarrow 0$  when we refine the partition ($\Box\rightarrow \cdot$).\\

More generally, we regularize the integral by  using a partition ${\cal P}^\epsilon$ which consists of:
\begin{itemize}
\item an $\epsilon$-dependent cellular decomposition ${\cal C}^\epsilon$ of $\Sigma$;
\item assigned to each cell $\Delta \in {\cal C}^\epsilon$:
\begin{itemize}
\item a point $x_\Delta$ inside $\Delta $;
\item a family of 2-surfaces $S^I_\Delta \subset \partial\Delta$, $I=1,...,n_\Delta$;
\item a family of paths $p_\Delta(x)$ labeled by points $x \in \partial\Delta$, each going from $x_\Delta$ to $x$;
\item a family of loops $\alpha_\Delta^K$, $K=1,...,m_\Delta$;
\item a family of coefficients $\kappa_{\Delta IJK}$;
\end{itemize} 
\end{itemize}
such that the following functional  
\be
 \int d^3 x f H^{\rm Eucl {\cal P}_\epsilon}(A,E)\ :=\ 
\frac{1}{\sqrt{8\pi G \beta^2}} \sum_{\Delta}  f(x_\Delta)\sqrt{\epsilon_{ijk} \sum_{IJK}\kappa_{\Delta IJK} P_{{S^I_\Delta},i} P_{{S^J_\Delta},j} (h_{\alpha^K_\Delta }^k)^{(l)}}
\ee    
approaches the Euclidean Hamiltonian,
\be
\int d^3 x f H^{\rm Eucl {\cal P}_\epsilon}(A,E)\ \underset{\epsilon\rightarrow 0}{\longrightarrow}\  \int d^3 x f H^{\rm Eucl}\ .
\ee
As in the cubic example, by $P_{S^I_\Delta,i}$ we mean $P_{S,\xi}$ of (\ref{flux-clas}) with 
\be
S=S^I_\Delta, \ \ \ {\rm and}\ \ \  \xi(x)\ :=\ h_{p_\Delta(x)}\tau_ih_{p_\Delta(x)}^{-1}\ .
\ee

Each term $\epsilon_{ijk} \kappa_{\Delta IJK} P_{{S^I_\Delta} i} P_{{S^J_\Delta} j} (h_{\alpha^K_\Delta }^k)^{(l)}$ gives rise to a well defined operator in ${\cal H}_{\rm kin}$ 
 \be\label{hPP}
 \epsilon_{ijk} \kappa_{\Delta IJK}\vdots\hat{P}_{{S^I_\Delta},i} \hat{P}_{{S^J_\Delta},j}(\hat{h}_{\alpha^K_\Delta }^k)^{(l)}\vdots
 \ee
which depends on the ordering of the operators, symbolized by $\vdots\quad\vdots$.

In this way we obtain an operator  
\be\label{heps}
\widehat{ \int d^3 x f(x) H^{\rm Eucl {\cal P}_\epsilon}(x)}\ :=\ 
\frac{1}{\sqrt{8\pi G \beta^2}} \sum_{\Delta}  f(x_\Delta)\sqrt{\epsilon_{ijk} \kappa_{\Delta IJK}\vdots\hat{P}_{{S^I_\Delta},i} \hat{P}_{{S^J_\Delta},j}(\hat{h}_{\alpha^K_\Delta }^k)^{(l)}\vdots}
 \ee   
which depends on the partition ${\cal P}_\epsilon$ and is well defined in ${\cal H}_{\rm kin}$. However, as we refine the partition ${\cal P}_\epsilon$, the operator family does not converge to any operator in ${\cal H}_{\rm kin}$. This is a well known problem in LQG and it does not have a solution in the kinematical Hilbert space ${\cal H}_{\rm kin}$. 

A way out is to consider the dual action of the regulated operators $\int d^3 x f(x) \hat{H}^{\rm Eucl {\cal P}_\epsilon}(x)$ in the Hilbert space ${\cal H}_{\rm vtx}^G$. That was done for the (formally regularized) operator $\hat{C}_{\rm gr}$ in \cite{Lewandowski:2014hza}. As it is explained therein, and those arguments apply also in the case at hand, a limit as $\epsilon\rightarrow 0$ exists upon several conditions about the partitions ${\cal P}_\epsilon$. To begin with, we adjust the partitions individually to each subspace ${\cal H}_\gamma$ in the decomposition (\ref{decomp}). Secondly, a successful partition has to have a suitable diffeomorphism covariance in the dependence of the partitions on $\gamma$ and on $\epsilon$.  

The outstanding problem though, is the dependence of the result on choices made. There are many partitions which satisfy the conditions. The resulting operator carries a memory of the choice of ${\cal P}_\epsilon$, for example on the adjustment of the fluxes to graphs. To restrict that ambiguity, we study first the straightforward quantization of $\epsilon_{ijk} F_{ab} E^a_i E^b_j$.

Let $\Psi\in {\rm Cyl}$ be as in (\ref{cyl'}). Assuming $\hat{E}^a_i(x)=\frac{\hbar}{i}\frac{\delta}{\delta A^i_a(x)}$, we obtain
\begin{align}
&\epsilon_{ijk}F_{ab}^k(x) \biggl(\frac{\hbar}{i}\biggr)^2 \frac{\delta}{\delta A_a^i(x)}\frac{\delta}{\delta A_b^j(x)}\psi(h_{e_1}(A),...,h_{e_n}(A))\ 
=\nonumber\\
&\sum_{e_I,e_{I'}}\int_{t_0}^{t_1}\int _{t'_0}^{t'_1} dt dt' \delta(x,e_I(t))\delta(x,e_{I'}(t')) F^k_{ab}(x)\dot e_I^a(t)\dot e_{I'}^b(t')\nonumber\\ 
&\left(h_{e_I,t_1,t}(A)\tau^ih_{e_{I},t,t_0}(A)\right)^A_B\left(h_{e_{I'},t'_1,t'}(A)\tau^jh_{e_{I'},t',t'_0}(A)\right)^{A'}_{B'}
\frac{\partial}{\partial g_I{}^A_B}  \frac{\partial}{\partial g_{I'}{}^{A'}_{B'}}\psi(g_1,...,g_n)_{|_{g_I=h_{e_I}(A), g_{I'}=h_{e_{I'}}(A)}}
\end{align} 
Certainly  the product of the two Dirac delta distributions is ill defined at some points $x$, $e_I(t)$ and $e_{I'}(t')$. However, we can precisely indicate those points at which the expression is identically zero. To begin with, the product $\delta(x,e_I(t))\delta(x,e_{I'}(t')) $ vanishes except for the triples $( x, e_I(t), e_{I'}(t') )$ such that
\be x=e_I(t)=e_{I'}(t') . \ee            
Secondly, the factor $F^k_{ab}(x)\dot e_I^a(t)\dot e_{I'}^b(t')$ is not zero only if 
\be \dot e_I(t) \nparallel \dot e_{I'}(t') . \ee
 
Those two conditions are satisfied simultaneously only if $x$ coincides with one of the vertices $v$ of $\gamma$ and the edges $e_I$ and $e_{I'}$ intersect transversally at $v$. Suppose that $x=v\in$Vert$(\gamma)$, the edges $e_I:[t_0,t_1]\rightarrow \Sigma$
and $e_{I'}:[t'_0,t'_1]\rightarrow \Sigma$ intersect transversally at $v$, both oriented to be outgoing. Then the corresponding contribution comes only from
\be x=v,\ \ \ \ t_I=t_0,\ \ \ \  t_{I'}=t'_0, \ee
and it is 
\be 
 F^k_{ab}(v)\dot e_I^a(t_)\dot e_{I'}^b(t'_0)
\left( h_{e_I}(A)\tau^i \right)^A_B \left( h_{e_{I'}}(A)\tau^j\right)^{A'}_{B'}
\frac{\partial}{\partial g_I{}^A_B}  \frac{\partial}{\partial g_{I'}{}^{A'}_{B'}}\psi(g_1,...,g_n)\bigg|_{g_I=h_{e_I}(A), g_{I'}=h_{e_{I'}}(A)}
\ee
modulo the ill defined factor $\left(\delta(v,v)\right)^2$ which has to be regularized. Our regularization is also expected to replace $F^k_{ab}\dot e_I \dot e_{I'}$ by a parallel transport $h_{e_{II'}}$ along a loop $e_{II'}$ assigned to the two (segments of) edges. Finally, diffeomorphism invariance implies that each vertex $v$ and a pair of transversally intersecting edges $e_I$ and $e_{I'}$ at $v$ contribute the same operator as any other diffeomorphism equivalent triple $v'$, $e'_I$ and $e'_{I'}$.    

We are now in a position to formulate assumptions about the construction of the partitions ${\cal P}_\epsilon$ adapted to a graph $\gamma$, as shown in \cite{Lewandowski:2014hza}, and the assumptions about the assignment of the loop $\alpha_\Delta^K$ used to regularize the connexion curvature $F^k_{ab}$, in order to guarantee the diffeomorphism covariance of the final operator.

Given a graph $\gamma=(e_1,...,e_n)$ of Vert$(\gamma)=(v_1,...,v_m)$, in order to spell out the conditions it is convenient to split each edge into two segments and orient the new edges to be outgoing from the vertices of the original graph $\gamma$. Denote the  resulting graph by $\gamma'=(e'_1,...,e'_{2n})$ and its vertex set Vert$(\gamma')=(v_1,...,v_m,v'_{m+1},...,v'_{m+n})$. The assumptions are as follows:  
\begin{req}\label{Requirement1}\ 

\begin{itemize}
\item each cell $\Delta$ contains at most one vertex of the graph $\gamma'$;
\item if $v \in $Vert$(\gamma)$ and $v\in\Delta$, then 
\begin{itemize}
\item $x_\Delta=v$;
\item to each edge $e'_I$ there is assigned a surface  $S^I_\Delta \subset \partial \Delta$ intersecting the edge transversally (there may be surfaces in $\partial \Delta$ not intersecting any edge);
\item to each ordered pair of edges $e'_I$ and $e'_J$  meeting transversally at $v$ there is assigned a loop $\alpha^{IJ}_\Delta$ oriented according to the order of the pair $(e'_I,e'_J)$. Hence we denote $\kappa_{\Delta IJK}$ as $\kappa_{\Delta IJIJ}$;
\item  for edges  $e'_I$ and $e'_J$ of $\gamma'$ meeting transversally at $v$ $\kappa_{\Delta IJIJ}$ is not zero;
\item for edges  $e'_I$ and $e'_J$ of $\gamma'$ meeting tangentially at $v$ $\kappa_{\Delta IJIJ}=0$;
\item for edges  $e'_I$ and $e'_J$ of $\gamma'$ meeting transversally at $v$ the corresponding loop $\alpha_{IJ}$, in the limit $\epsilon\rightarrow 0$, is shrank to $v$ in a diffeomorphism invariant way;
\item if $\Delta$ does not contain an edge of $\gamma$ but it contains a segment of an edge then, by splitting the edge and reorienting its segments suitably, we turn that case into the case of $\Delta$ containing a 2-valent vertex;
\end{itemize}
\item the value of non vanishing $\kappa_{\Delta IJIJ}$ is an overall constant $\kappa_1(v)$ depending on the valence of the vertex but independent of $\Delta, I, J$.
\end{itemize}
\end{req}

Concerning the prescription for the assignment of the loops $\alpha_{IJ}$ -- we call them {\it special loops} -- which are created by the Euclidean part of our Hamiltonian operator, we wish the construction to satisfy the following requirements:
\begin{itemize}
\item[--] The loop added by the Hamiltonian should be attached to the graph according to a diffeomorphism invariant prescription
\cite{lqgcan1,Lewandowski:2014hza}. This property allows the operator to be well defined on the space ${\cal H}_\text{vtx}^G$.
\item[--] It should be possible to distinguish between loops attached to the same vertex but associated to different pair of edges, and between loops attached to the same pair of edges by successive actions of the Hamiltonian. This property makes it possible to define the adjoint operator on a dense domain in ${\cal H}_\text{vtx}^G$, and consequently to construct a symmetric Hamiltonian operator.
\end{itemize}

Consider a vertex $v$ of the graph $\gamma$ defined above and a set of links $\{e_I\}$ incident at $v$. In order to satisfy the first requirement, we use a construction that was introduced in \cite{Diff.Pr.} and was presented in a work of T. Thiemann \cite{Thiemann96a}. The construction consists of two parts. Firstly, to each pair of links $e_I$ and $e_J$ incident at $v$, we define an adapted frame in a small enough neighborhood of $v$. Then we require that the loop $\alpha_{IJ}$, associated to the pair $(e_I,e_J)$, lies in the coordinate plane spanned by the edges $e_I$ and $e_J$. The choice of the adapted frame is based on the following lemma:

\begin{quote}
{Let $e$ and $e'$ be two distinct analytic curves intersecting only at their starting point $v$. Then there exist parameterizations of these curves, a number $\delta > 0$, and an analytic diffeomorphism such that, in the corresponding frame, the curves are given by
\begin{itemize}
\item[(a)] $e (t) = (t, 0, 0)$, $e' (t) = (0, t, 0)$, $t \in [0, \delta]$ if their tangents are linearly independent at $v$,
\item[(b)] $e (t) = (t, 0, 0)$, $e' (t) = (t, t^n , 0)$, $t \in [0, \delta]$ for some $n \geq 2$ if their tangents are co-linear at $v$.
\end{itemize}
We will call the associated frame a frame adapted to $e$, $e'$.}
\end{quote}

To carry out the second part of the construction, we need a diffeomorphism invariant prescription of the topology of the routing of the loop $\alpha_{IJ}$. In other words, the plane in which the loop lies should be chosen in a way which is diffeomorphism invariant, and which does not cause the loop to intersect the graph $\gamma$ at any point different from the vertex $v$. The choice that $\alpha_{IJ}$ lies in a small enough neighborhood of $v$ guarantees that the loop cannot intersect any edge of $\gamma$ except the edges incident at the vertex $v$. Then the routing of the loop in that neighborhood is achieved through the prescription given in \cite{Thiemann96a} (and which we do not repeat here).

Now let us turn to the second requirement, which is crucial in order to have the possibility of defining a dense adjoint operator that allows one to construct symmetric Hamiltonian operators, and eventually to provide self-adjoint extensions. To state the prescription that satisfies the second requirement, we need to define the {\em order of tangentiality} of an edge at the node. This is defined as follows. Considering the vertex $v$ and the edge $e_I$, we denote by $k_{IJ}\geq 0$ the order of tangentiality of $e_I$ with another edge $e_J$ incident at $v$. If the edges $e_I$ and $e_J$ are not tangent at $v$, we understand that $k_{IJ}=0$. The order of tangentiality $k_I$ of the edge $e_I$ at the vertex $v$ 
\be
k_I = \max_{\substack{\text{$e_J$ at $v$} \\ J\neq I}} k_{IJ}
\ee
i.e. as the highest order of tangentiality of the edge $e_I$ with the remaining edges incident at $v$.

The element which completes the prescription of the special loop according to the two requirements is now stated as follows:
\begin{req}\label{Requirement2}\ 

\begin{quote}
{ The special loop $\alpha_{IJ}$ is tangent to the two edges $e_I$ and $e_J$ at the vertex $v$ up to orders $k_I + 1$ and $k_J + 1$ respectively, where $k_I (\geq 0)$ and $k_J (\geq 0)$ are respectively the orders of tangentiality of $e_I$ and $e_J$ at the node.}
\end{quote}
\end{req}
This property indeed makes a loop attached by the Hamiltonian to a given pair of edges perfectly distinguishable from any other loop at the same node.

To summarize, the prescription for assigning a special loop to a pair of links incident at a vertex is to choose the loop to lie in the coordinate plane defined by the frame adapted to the pair of edges, then to follow a specific and well defined routing of the loop described in \cite{Thiemann96a}, and finally to impose the tangentiality conditions introduced above. With this prescription, the loop assigned to a pair of edges is unique up to diffeomorphisms.\\
 
In consequence, given a graph $\gamma$ and the auxiliary graph $\gamma'$ obtained by the splitting, the contribution from a cell $\Delta$ containing a vertex $v$ reads
\be
\kappa_1(v)  \sum_{I,J} \epsilon_{ijk}\ \epsilon\left(\dot{e}'_I,\dot{e}'_J\right)\vdots\hat{J}_{v,e'_I,i} \hat{J}_{v,e'_J,j}(\hat{h}^k_{\alpha_{IJ}^{(\epsilon)}})^{(l)}\vdots \ \ =:\ \widehat{H_v^E}^{(\epsilon)}
\ee   
where $\epsilon\left(\dot{e}'_I,\dot{e}'_J\right)$ is $0$ if $\dot{e}'_I$ and $\dot{e}'_J$ are linearly dependent or $1$ otherwise. This operator maps
\begin{align} 
\widehat{H_v^E}^{(\epsilon)}:{\rm Cyl}_\gamma\ \rightarrow\  {\rm Cyl}_{\gamma''}\nonumber\\
\gamma''\ =\ \gamma \bigcup_{IJ}\{\alpha_{IJ}\}.
\end{align}
Considering all the graphs we combine the operators into a single $\epsilon$-dependent operator
\be\label{H-euc-epsilon}  \widehat{H_v^E}^{(\epsilon)} : {\rm Cyl}\rightarrow {\rm Cyl}\ .\ee

In order for the operator \eqref{H-euc-epsilon} to be cylindrically consistent, we should have $\kappa_1(v) = \kappa_1$, an overall constant independent of the valence of the vertex $v$.\footnote{On the other hand, every value of the constant $\kappa_1$ can be achieved by a suitable choice of the shape and size of the loops $\alpha_{IJ}$.} However, since our goal at the end is to implement this operator in the gauge invariant Hilbert space, we can equally well define the operator by proceeding with the regularization 
directly on the spaces Cyl$_\gamma^G$ orthogonal to each other. In that case the question of cylindrical consistency does not arise, and we may allow the possibility that $\kappa_1(v)$ depends on the valence of the vertex.

In this way we have determined the action of an operator (\ref{CF}) up to a value of $\kappa_1(v)$ (constant or not), assuming the conditions (\ref{Requirement1}) and (\ref{Requirement2}). This operator passes naturally to ${\rm Cyl}^G$
\be  \widehat{H_v^E}^{(\epsilon)} : {\rm Cyl}^G\rightarrow {\rm Cyl}^G\ .\ee

As we refine the partition by $\epsilon\rightarrow 0$, the loops $\alpha_{IJ}$ are shrank to $v$. However, the $\epsilon$-dependent operator $\left(\widehat{H_v^E}^{(\epsilon)}\right)^*$ defined by the duality ${}^*$ in ${\cal H}_{\rm vtx}^G$ (on a domain that includes $\eta({\rm Cyl^G})$),
\begin{align}\label{EucDual}
\eta(\Psi) &\mapsto \left(\widehat{H_v^E}^{(\epsilon)}\right)^*\eta(\Psi)\nonumber\\
 \left(\widehat{H_v^E}^{(\epsilon)}\right)^*\eta(\Psi) : \Psi'&\mapsto \eta(\Psi)(\widehat{H_v^E} \Psi') .
 \end{align}
is insensitive to the shrinking, as long as each loop $\alpha_{IJ}$ is shrank within the diffeomorphism class of $\gamma \cup \alpha_{IJ}$. Hence we drop the $\epsilon$ label in the dual operator. It follows that the Euclidean part of the Hamiltonian is defined as
\be\label{heps2}
 \widehat{\int_\Sigma d^3 x f(x) H^{\rm Eucl}(x)}\ :=\ 
\frac{1}{\sqrt{8\pi G \beta^2}} \sum_{v \in \Sigma}  f(v)\sqrt{\widehat{H_v^E}^*}
 \ee   
In order to define the square root in this equation, one could choose a symmetric ordering of $\widehat{H^E}^*_v$. However, a symmetric ordering of the Euclidean term is not necessary for constructing the complete Hamiltonian, for which instead the square root of the sum of Euclidean and Lorentzian terms needs to be defined.
 
\subsubsection{Lorentzian part}

Following the strategy of quantization indicated by equation \eqref{reg A2+B2}, we now introduce a second operator corresponding to the integral of the term $\sqrt{qR}$ again smeared with an arbitrary function $f$
$$ \int d^3x f(x) \sqrt{qR}(x) . $$
The construction of the operator is in two parts: first we write an approximate expression of the classical integral by implementing a cellular decomposition $\mathscr{C^\epsilon}$ of the 3d manifold, characterized by a regulator $\epsilon$. Secondly, the regularized expression is promoted to an operator, which after taking the regulator limit, leads to a background independent operator acting in the Hilbert space of gauge invariant states. 

The aim is to construct an operator corresponding to the following function on the classical phase space:
\begin{align}\label{Loren_part1}
\int_\Sigma d^3x\ f(x)\sqrt{\abs{\det[E]}\ R}=\int_\Sigma d^3x\ f(x)\sqrt{\sqrt{\abs{\det[E]}}\ \sqrt{\abs{\det[E]}} R},
\end{align}

Consider a cellular decomposition $\mathscr{C^\epsilon}$ of the manifold $\Sigma$. The size of the cells is assumed to be controlled by the regulator $\epsilon$, in such a way that the coordinate size $\epsilon_{\Delta'}$ of each cell $\Delta'\in \mathscr{C^\epsilon}$ satisfies $\epsilon_{\Delta'}<\epsilon$. We can then write the integral \eqref{Loren_part1} as a limit of a Riemannian sum over the cells $\Delta'$,
\be
\int_\Sigma d^3x\ f(x)\sqrt{\sqrt{\abs{\det[E]}}\ \sqrt{\abs{\det[E]}} R} = \lim_{\epsilon \rightarrow 0} \sum \limits_{\Delta' \in \mathscr{C^\epsilon}} f(x_{\Delta'})\sqrt{\biggl(\int_{\Delta'} d^3x\,\sqrt{\abs{\det[E]}}\biggr)\biggl(\int_{\Delta'} d^3x\,\sqrt{\abs{\det[E]}}R\biggr)}, \label{Lorentz-step1}
\ee
where on the right-hand side $x_{\Delta'}$ denotes any point inside $\Delta'$.

Next we decompose each cell $\Delta'$ into $c_{\Delta'}$ closed cells $\Delta$, where a cell $\Delta$ has a boundary formed by a number $n_\Delta$ of 2-surfaces (faces). In equation \eqref{Lorentz-step1}, we then approximate the integral of $\sqrt{\abs{\det[E]}}$ by a Riemannian sum over the cells $\Delta$, and the integral of $\sqrt{\abs{\det[E]}}R$ by a regularized Regge action for an appropriate $\Delta$-decomposition of $\Delta'$, obtaining
\be
\int_\Sigma d^3x\ f(x)\sqrt{\sqrt{\abs{\det[E]}}\ \sqrt{\abs{\det[E]}} R} = \lim_{\epsilon \rightarrow 0} \sum \limits_{{\Delta'} \in \mathscr{C^\epsilon}} f(x_{\Delta'})\sqrt{\left(\sum \limits_{\Delta \subset \Delta'} \sqrt{q_\Delta(E)}\right) \left(\sum \limits_{\Delta \subset \Delta'} R_\Delta(E)\right)}. \label{Lorentz-step2}
\ee
The functionals $q_\Delta(E)$ \cite{RovelliSmolin95} and $R_\Delta(E)$ are defined on the classical phase space as\footnote{The functional $q_\Delta(E)$ can be defined in a different way:
\begin{align}
 q_\Delta(E):=\sum \limits_{I J K}\abs{\frac{ \kappa_0(\Delta)}{3!} \epsilon_{IJK} \epsilon_{ijk} P_{S_\Delta^I,i} P_{S_\Delta^J,j} P_{S_\Delta^K,k}}.
\end{align}
This definition would lead to a volume operator that is sensitive to the differential structure at the nodes, see \cite{AshtekarLewand95, AshtekarLewand98, KristThiemann06}.
}
\begin{align}
q_\Delta(E):= \frac{ \kappa_0(\Delta)}{3!} \sum \limits_{I \neq J \neq K} \abs{\epsilon_{ijk} P_{S_\Delta^I,i} P_{S_\Delta^J,j} P_{S_\Delta^K,k}},
\end{align}

\begin{align}\label{reg_L}
R_\Delta(E):= \sum_{u \subset \partial \Delta}\ &\sqrt{\delta_{ii'} \frac{\frac{1}{2} \epsilon_{ijk} P_{S_{\Delta,u}^I,j} P_{S_{\Delta,u}^J,k}} {\sqrt{q_\Delta(E)}}\ \frac{\frac{1}{2} \epsilon_{i'j'k'} P_{S_{\Delta,u}^{I},j'} P_{S_{\Delta,u}^{J},k'}}{\sqrt{q_\Delta(E)}}} \\ \nonumber & \times \left(\frac{2\pi}{\alpha_u}-\pi + \arccos \left[ \frac{ \delta_{kl} P_{S_{\Delta,u}^I,k} P_{S_{\Delta,u}^J,l}}{2|P_{S_{\Delta,u}^I}||P_{S_{\Delta,u}^J}|} \right] \right),
\end{align}
where we use the following notation:
\begin{itemize}
\item[--] given $\Delta$, the index $I=1,...,n_\Delta$ labels the surfaces (faces) $S_\Delta^I$ forming the boundary $\partial \Delta$  of the cell $\Delta$ and $u$ labels the hinges on that boundary (the 1-skeleton of the cell); 
\item[--] the symbols  $S_{\Delta,u}^I$ and  $S_{\Delta,u}^J$ stand for the two surfaces in $\partial \Delta$ that intersect at $u$;
\item[--] the symbol  $P_{S_\Delta^I,i}$ represents the flux of the field $E^a_i$ across $S_{\Delta}^I$, defined in (\ref{flux-clas}) with 
\be
S=S_\Delta^I, \ \ \ {\rm and}\ \ \  \xi(x)\ :=\ h_{p_\Delta(x)}\tau_ih_{p_\Delta(x)}^{-1}
\ee
and
\be
|P_{S_{\Delta,u}^I}|:=\sqrt{\delta_{kk'}P_{S_{\Delta,u}^I,k}P_{S_{\Delta,u}^I,k'}};
\ee

\item[--] $\kappa_0(\Delta)$ is a regularization constant depending on the shape of the cell $\Delta$;

\item[--] finally $\alpha_u$ is a fixed integer parameter corresponding to the number of cells sharing the hinge $u$ in the cellular decomposition $\mathscr{C^\epsilon}$.
\end{itemize}

Considering the coordinate size $\epsilon_{\Delta}<\epsilon$ of the cell $\Delta$, defined such that the limit $\epsilon\to 0$ is equivalent to $\epsilon_{\Delta}\to 0$, the functional $q_\Delta(E)$ is such that $\frac{1}{V_\Delta^2}q_\Delta(E)$ approximate the function $\abs{\det[E]}$ at any point within the cell $\Delta$, $V_\Delta\propto \epsilon_{\Delta}^3$ being the coordinate volume of $\Delta$. 
Also, each term in the sum defining $R_\Delta(E)$ (\ref{reg_L}), rescaled by $L_u\propto \epsilon_{\Delta}$ that is the coordinate length of the edge $u$ on the boundary of $\Delta$,
\begin{align}
&\frac{1}{L_u} \sqrt{\delta_{ii'} \frac{\frac{1}{2} \epsilon_{ijk} P_{S_{\Delta,u}^I,j} P_{S_{\Delta,u}^J,k}} {\sqrt{q_\Delta(E)}}\ \frac{\frac{1}{2} \epsilon_{i'j'k'} P_{S_{\Delta,u}^{I},j'} P_{S_{\Delta,u}^{J},k'}}{\sqrt{q_\Delta(E)}}} \left(\frac{2\pi}{\alpha_u}-\pi + \arccos \left[ \frac{ \delta_{kl} P_{S_{\Delta,u}^I,k} P_{S_{\Delta,u}^J,l}}{2|P_{S_{\Delta,u}^I}||P_{S_{\Delta,u}^J}|} \right] \right),
\end{align}
approximate the function $L_u(E) \Theta_u(E)$ in the limit $\epsilon_{\Delta}\to 0$, where $L_u(E)$ and $\Theta_u(E)$ are respectively the length of the hinge $u$ and the dihedral angle at $u$ in $\Delta$ expressed in terms of densitized triads.

The sum over the cells $\Delta$ of the functional $R_\Delta(E)$ corresponds to the regularized Regge action \cite{Regge1} in $3d$ on $\Delta'$, which is by itself an approximation of the function $\int_{\Delta'} d^3x\ \sqrt{\abs{\det[E]}} R$. We direct the reader to \cite{Curvature_op.} for more details about the concepts of this construction.\\

To continue the calculation from equation \eqref{Lorentz-step2}, we assume that the cells $\Delta$ are chosen such that we obtain the same contributions $q_\Delta(E)$ and $R_\Delta(E)$ from each cell $\Delta$, up to higher order corrections in $\epsilon_{\Delta'}$ (equivalently, up to higher order corrections in $\epsilon$). Hence each sum over the cells $\Delta$ becomes the number of cells $c_{\Delta'}$ times the contribution of the cell $\tilde{\Delta}$, chosen as the cell containing the point $x_{\Delta'}$ at which the smearing function $f$ is evaluated. In this way we obtain
\be\label{Lorentz-step3}
\int_\Sigma d^3x\ f(x)\sqrt{\sqrt{\abs{\det[E]}}\ \sqrt{\abs{\det[E]}} R} = \lim_{\epsilon \rightarrow 0} \sum \limits_{{\Delta'} \in \mathscr{C^\epsilon}} f(x_{\tilde{\Delta} \subset \Delta'})\  c_{\Delta'} \sqrt{\sqrt{q_{\tilde{\Delta}}(E)}\ R_{\tilde{\Delta}}(E)}.
\ee

Let us now introduce the approximation
\be\label{f-average}
f(x_{\tilde{\Delta} \subset \Delta'}) = \frac{1}{c_{\Delta'}} \sum \limits_{\Delta \subset \Delta'} f(x_{\Delta}),
\ee
which is an averaging of the values of the function $f$ inside the cell $\Delta'$, and which can be seen as a better approximation of the value of the function $f$ inside the cell $\Delta'$, in the sense that we are probing the function $f$ in several points inside the cell instead of one point $x_{\Delta'}$. Inserting equation \eqref{f-average} in equation \eqref{Lorentz-step3}, we come to the result we are looking for:
\begin{align}
\int_\Sigma d^3x\ f(x)\sqrt{\sqrt{\abs{\det[E]}}\ \sqrt{\abs{\det[E]}} R} &= \lim_{\epsilon \rightarrow 0} \sum \limits_{{\Delta'} \in \mathscr{C^\epsilon}} \sum \limits_{\Delta \subset \Delta'} f(x_{\Delta}) \sqrt{\sqrt{q_{\Delta}(E)}\ R_{\Delta}(E)} \notag \\
&= \lim_{\epsilon \rightarrow 0} \sum \limits_{\Delta \in \mathscr{C^\epsilon}} f(x_{\Delta}) \sqrt{\sqrt{q_{\Delta}(E)}\ R_{\Delta}(E)}, 
\label{Loren_reg}
\end{align}
where last step is achieved by combining the two sums over $\Delta'$ and $\Delta$.\\

Notice that the expression of $R_\Delta(E)$ in (\ref{reg_L}) contains an overall factor of $\bigl(\sqrt{q_\Delta(E)}\bigr)^{-1}$. This leads to a crucial simplification in the expression of $\sqrt{q_\Delta(E)}\ R_\Delta(E)$, namely, the factors of $\sqrt{q_\Delta(E)}$ are canceled:
\begin{align}\label{Loren_ex}
\sqrt{q_\Delta(E)}\ R_\Delta(E)=\sum_{u \subset \partial \Delta}\ \frac{1}{2} &\sqrt{ \delta_{ii'} \epsilon_{ijk} P_{S_{\Delta,u}^I,j} P_{S_{\Delta,u}^J,k}\ \epsilon_{i'j'k'} P_{S_{\Delta,u}^{I},j'} P_{S_{\Delta,u}^{J},k'}} \\ \nonumber & \times \left(\frac{2\pi}{\alpha_u}-\pi + \arccos \left[ \frac{ \delta_{kl} P_{S_{\Delta,u}^I,k} P_{S_{\Delta,u}^J,l}}{2|P_{S_{\Delta,u}^I}||P_{S_{\Delta,u}^J}|} \right] \right),
\end{align}
In the quantum theory, this simplification implies that the volume operator will be absent from the Lorentzian part, and consequently from the whole Hamiltonian operator. The absence of the volume is an important technical advantage in the calculation of the action of the Hamiltonian.\\

Before promoting this expression to an operator, we study the term $\epsilon_{ijk} P_{S_{\Delta,u}^I,j} P_{S_{\Delta,u}^J,k}$ appearing in equation (\ref{Loren_ex}). This term approximates the classical function
\begin{align}
 \epsilon_{ijk} \epsilon_{abc} E_j^a E_k^b \dot{u}^c(s) =\lim \limits_{\epsilon_\Delta \rightarrow 0} \frac{1}{\epsilon_\Delta^{\ 4}}\epsilon_{ijk} P_{S_{\Delta,u}^I,j} P_{S_{\Delta,u}^J,k},
\end{align}
where $s$ is parameterizing the curve $u$.\\

Considering $\Psi$ in $\text{Cyl}$ with a graph $\gamma=(e_1,...,e_n)$, the straightforward quantization of $\epsilon_{ijk} \epsilon_{abc} E_j^a E_k^b \dot{u}^c(s)$ by replacing $E_j^a$ with $\frac{\hbar}{i}\frac{\delta}{\delta A^j_a(x)}$ induces the factor
\begin{align}
\epsilon_{abc} \dot{e}_I^a(t) \dot{e}_J^b(t') 
\end{align}
in the formal action of the operator on $\Psi$. Similarly to the case of the Euclidean part (see section \ref{EuclideanPart}), this factor vanishes unless 
\be \dot e_I(t) \nparallel \dot e_{J}(t') . \ee
which means that the edges $e_I(t)$ and $e_J(t')$ are different ($I \neq J$) and transversal at their intersection point. In order to pass this property to the quantum operator, we introduce the coefficient $\kappa'_{\Delta IJ}$, defined in the following, in the expression of $\sqrt{q_\Delta(E)}\ R_\Delta(E)$ and we write
\begin{align}\label{Loren_exp}
\sqrt{q_\Delta(E)}\ R_\Delta(E) :=\sum_{u \subset \partial \Delta}\ \kappa'_{\Delta IJ}\ \frac{1}{2} &\sqrt{ \delta_{ii'} \epsilon_{ijk} P_{S_{\Delta,u}^I,j} P_{S_{\Delta,u}^J,k}\ \epsilon_{i'j'k'} P_{S_{\Delta,u}^{I},j'} P_{S_{\Delta,u}^{J},k'}} \\ \nonumber & \times \left(\frac{2\pi}{\alpha_u}-\pi + \arccos \left[ \frac{ \delta_{kl} P_{S_{\Delta,u}^I,k} P_{S_{\Delta,u}^J,l}}{2|P_{S_{\Delta,u}^I}||P_{S_{\Delta,u}^J}|} \right] \right),
\end{align}

In order to promote the expression in (\ref{Loren_exp}) to a quantum operator, we first need to set some requirements on the decomposition $\mathscr{C_\Sigma^\epsilon}$ so that we adapt it to the functions in $\text{Cyl}$: given a $\Psi$ in $\text{Cyl}$ with a graph $\gamma=(e_1,...,e_n)$ of Vert$(\gamma)=(v_1,...,v_m)$, the requirements are as follows:
{\it
\begin{itemize}
\item each cell $\Delta$ contains at most one vertex of the graph $\gamma$;
\item each 2-cell (face) on the boundary of a cell $\Delta$, containing a vertex of $\gamma$, is punctured exactly by one edge of the graph $\gamma$. The intersection is transversal and belongs to the interior of the edge;
\item if $v \in $Vert$(\gamma)$ and $v \in \Delta$, then 
\begin{itemize}
\item[--] $x_\Delta=v$,
\item[--] $\kappa'_{\Delta IJ}$ is not zero only for edges $e_I$ and $e_J$ of $\gamma$ meeting transversally at $v$;
\end{itemize}
\item if $\Delta$ does not contain an edge of $\gamma$ but it contains a segment of an edge then, by splitting the edge and reorienting its segments suitably, we turn that case into the case of $\Delta$ containing a 2-valent vertex,
\end{itemize}
}

A result that follows from the derivation of the curvature operator in \cite{Curvature_op.} is that the value of non vanishing $\kappa'_{\Delta IJ}$ is an overall constant $\kappa_2(v)$ depending on the valence of the vertex but independent of $\Delta, I, J$. This property is obtained from an averaging procedure used in order to remove the dependence on $\mathscr{C_\Sigma^\epsilon}$.\\

Having the quantum operators corresponding to $P_{S_{\Delta,u}^I,j}$ s', we are now able to define the quantum operator corresponding to ${q_\Delta(E)\ R_\Delta(E)}$ as
\begin{align}\label{Loren_part3}
 \widehat{q_\Delta(E)\ R_\Delta(E)}:=\widehat{H_\Delta^L}:=\ \sum_{u \subset \partial \Delta}\ &\kappa'_{\Delta IJ}\ \frac{1}{2} \sqrt{ \delta_{ii'} \epsilon_{ijk} \hat{P}_{S_{\Delta,u}^I,j} \hat{P}_{S_{\Delta,u}^J,k}\ \epsilon_{i'j'k'} \hat{P}_{S_{\Delta,u}^{I},j'} \hat{P}_{S_{\Delta,u}^{J},k'}} \\ \nonumber & \times \left(\frac{2\pi}{\alpha_u}-\pi + \arccos \left[ \frac{ \delta_{kl} \hat{P}_{S_{\Delta,u}^I,k} \hat{P}_{S_{\Delta,u}^J,l}}{2\delta_{kk'} \hat{P}_{S_{\Delta,u}^I,k}\hat{P}_{S_{\Delta,u}^I,k'}\delta_{ll'} \hat{P}_{S_{\Delta,u}^J,l}\hat{P}_{S_{\Delta,u}^J,l'}} \right] \right).
\end{align}

Considering a cylindrical function $\Psi_\gamma$ in the Hilbert space $\text{Cyl}_{\gamma}$, thanks to regularization detailed above we have
\begin{align}\label{Loren_op1}
 \widehat{H_\Delta^L} \ \Psi_\gamma= \left\{
  \begin{matrix}
    \widehat{H_v^L} \Psi_{\gamma}\quad &\text{if $\Delta$ contains a vertex $v$ of $\gamma$;} \\
    0 &\text{if $\Delta$ does not contain a vertex $v$ of $\gamma$.}
  \end{matrix}
 \right.
\end{align}
where
\begin{align}\label{Curv1}
\widehat{H_v^L}:\ &=\ \kappa_2(v)  \sum \limits_{I,J}  \epsilon\left(\dot{e}_I,\dot{e}_J\right)\ \widehat{H_v^L}_{e_I,e_J} \\ \nonumber &=\ \kappa_2(v)  \sum \limits_{I,J}  \frac{\epsilon\left(\dot{e}_I,\dot{e}_J\right)}{2} \sqrt{\delta_{ii'} (\epsilon_{ijk} \hat{J}_{v,e_I,j} \hat{J}_{v,e_J,k}) (\epsilon_{i' j' k'} \hat{J}_{v,e_I,j'} \hat{J}_{v,e_J,k'})} \\ \nonumber
&\qquad \qquad \quad \left(\frac{2\pi}{\alpha_{v,e_I,e_J}} - \pi + \arccos \left[ \frac{ \delta_{kl} \hat{J}_{v,e_I,k} \hat{J}_{v,e_J,l}}{\sqrt{\delta_{kk'} \hat{J}_{v,e_I,k} \hat{J}_{v,e_I,k'}} \sqrt{\delta_{ll'} \hat{J}_{v,e_J,l} \hat{J}_{v,e_J,l'}}} \right]\right),
\end{align}
with $\alpha_{v,e_I,e_J}$ an integer parameter, $\kappa_2(v)$ is the averaging coefficient that depends only on the valence of the vertex $v$, and $\epsilon\left(\dot{e}_I,\dot{e}_J\right)$ is $0$ if $\dot{e}_I$ and $\dot{e}_J$ are linearly dependent or $1$ otherwise.\\

Now we can define an operator acting in $\text{Cyl}^G$
\begin{align}\label{Loren_op2}
 \sqrt{\frac{(1+\beta^2)}{8\pi\beta^2G}} \widehat{\int_\Sigma d^3x\ f(x)\sqrt{\abs{\det[E]}\ R}}:=  \sqrt{\frac{(1+\beta^2)}{8\pi\beta^2G}} \sum \limits_{v \in \Sigma} f(v)\sqrt{\widehat{H_v^L}},
\end{align}
that corresponds to the Lorentzian part of the (smeared) Hamiltonian. Consequently we introduce the operator $\widehat{H_v^L}^*$ defined by duality ${}^*$ on ${\cal H}_{\rm vtx}^G$
\begin{align}\label{LorDual}
 \sqrt{\frac{(1+\beta^2)}{8\pi\beta^2G}}\widehat{\int_\Sigma d^3x\ f(x)\sqrt{\abs{\det[E]}\ R}}:=  \sqrt{\frac{(1+\beta^2)}{8\pi\beta^2G}} \sum \limits_{v \in \Sigma} f(v)\sqrt{\widehat{H_v^L}^*},
\end{align}
Notice that there is no ordering ambiguity in the operator $\widehat{H_v^L}$ and therefore no ordering ambiguity in $\widehat{H_v^L}^*$.

\subsubsection{The Hamiltonian operator \& its properties}

At this level two operators $\widehat{H_x^E}^*$ and $\widehat{H_x^L}^*$ have been implemented on ${\cal H}_\text{vtx}$. Now we introduce the operator $\widehat{H_x^T}$ defined as
{\small
\begin{align}\label{Hamil1}
\widehat{H_x^T} &:= \widehat{H_x^E}^* + (1+\beta^2) \widehat{H_x^L}^* \notag \\
 & = \ \sum \limits_{e,e'}\epsilon\left(\dot{e},\dot{e}'\right) \left[ \left(\widehat{H_x^E}_{e,e'} \right)^*\ +\ (1+\beta^2) \left(\widehat{H_x^L}_{e,e'} \right)^*\right]
\end{align}
}
where $\dot{e}$, $\dot{e}'$ run through the set of germs of bounded one-dimensional sub-manifolds of $\Sigma$ incident to a point $x$, and we define $\widehat{H_x^E}_{e,e'}$ as
\begin{align}
\widehat{H_x^E}_{e,e'}\ :=\ \kappa_1(x) \epsilon_{ijk}\ (\hat{h}_{\alpha_{x,ee'}}^k)^{(l)} \hat J_{x,e,i} \hat J_{x,e',j}\ ,
\end{align}
by choosing the simplest ordering for the Euclidean part.\\
\vspace{0,3cm}

\paragraph{A symmetric Hamiltonian operator\\}
$ $\\
The Hamiltonian operator represents the quantization of the classical Hamiltonian of the deparametrized theory of GR coupled to a free scalar field. This final operator is required to be self-adjoint on some non-trivial domain in order for it to generate unitary evolution of the quantum system and for its spectra to admit a physical interpretation. Therefore, a first step toward achieving self-adjointness\footnote{The self-adjointness of the Hamiltonian operator presented in equation \eqref{Sym.Hamil} is still an open question. However, we are able to construct self-adjoint extensions using other symmetrizations.} of the Hamiltonian is to construct a symmetric operator.

A symmetric operator $\hat{H}$ could be introduced as a combination of the operator $\widehat{H_x^T}$ and its adjoint operator $\widehat{H_x^T}^\dagger$. The later is defined on a domain $\mathscr{D}\left(\widehat{H_x^T}^\dagger\right) \supset {\cal S}^G$, such that for every two states $\ket{\Psi}$ and $\ket{\Psi'}$ in the space ${\cal S}^G$ we have
\begin{equation}\label{adjoint}
 \bra{\Psi'}\widehat{H_x^T}^\dagger \ket{\Psi}= \overline{\bra{\Psi}\widehat{H_x^T} \ket{\Psi'}},
\end{equation}
where the bar stands for complex conjugate.

For instance we choose the following definition for $\hat{H}$ 
\begin{align}\label{Sym.Hamil}
\nonumber \hat{H} &:\mathscr{D}\left(\widehat{H_x^T}^\dagger\right) \cap \mathscr{D}\left(\widehat{H_x^T}\right) \supset {\cal S}^G\ \longrightarrow \ {\cal H}_\text{vtx}\\ 
 \hat{H} &:=\frac{1}{\sqrt{16\pi G \beta^2}} \sum \limits_{x\in \Sigma} \sqrt{\widehat{H_x^T}+\widehat{H_x^T}^\dagger}.
\end{align}

\vspace{0,3cm}

\paragraph{Gauge and diffeomorphism invariance\\}
$ $\\
The operators $\hat{h}_{\alpha_{ee'}}$ are $\hat J_{x,e,i}$ are both gauge invariant. Hence the Hamiltonian operator $\hat{H}$ is gauge invariant.

Considering the group of smooth diffeomorphisms, the operator $\hat{H}$ is also diffeomorphism invariant, thanks to the regularization adopted and the averaging procedures involved in defining the curvature operator \cite{Curvature_op.}. As a consequence of its gauge and diffeomorphism invariance, $\hat H$ preserves the space ${\cal H}_\text{vtx}^G$. \\
\vspace{0,3cm}

\paragraph{Action of the symmetric Hamiltonian operator\\}
$ $\\
Looking into the action of $\hat{H}$ on a spin network state $\ket{\gamma,\{j\},\{\iota\}}$, one would like to express the matrix elements of this operator in terms of the quantum numbers labeling the states, namely the spins $j$ and the intertwiners $\iota$. However in order to obtain such an expression, it is necessary to diagonalize the operator $\widehat{H_x^T}$ under the square root in the definition of $\hat{H}$. So far this aim of diagonalizing this operator has not been realized. Nevertheless, what can be achieved is the explicit computation of the action of $\widehat{H_x^T}$ on a spin network basis state $\ket{\gamma,\{j\},\{\iota\}}$. We need first to calculate the action of $\widehat{H_x^E} + (1+\beta^2) \widehat{H_x^L}$ then from (\ref{EucDual}), (\ref{LorDual}) and (\ref{adjoint}) we can deduce the action of the dual operators, hence the action of $\widehat{H_x^T}$ and its adjoint.

Moreover, having the state $\ket{\gamma,\{j\},\{\iota\}}$, we know from (\ref{Hamil1}) that the action of $\hat{H}$ reduces to a sum of contributions, each coming from a pair of edges incident to the same vertex $v$ of the graph $\gamma$; we call such a pair a {\it wedge}. The operators $\widehat{H_v^T}_{e,e'}$, each associated to such a wedge in $\gamma$, can be introduced as
{\small
\begin{align}\label{S.Hamil4}
\widehat{H_v^T}_{e,e'} \ket{\gamma,\{j\},\{\iota\}} &:= \epsilon\left(\dot{e},\dot{e}'\right) \left[ \left(\widehat{H_v^E}_{e,e'}\right)^*\ +\ (1+\beta^2) \left(\widehat{H_v^L}_{e,e'} \right)^*\right] \ket{\gamma,\{j\},\{\iota\}}.
\end{align}
}
Consequently we just need to compute the action of $\widehat{H_v^E}_{e,e'}$ and $\widehat{H_v^L}_{e,e'}$ leading to the total action of  $\widehat{H_v^T}_{e,e'}$ on $\ket{\gamma,\{j\},\{\iota\}}$. The result is presented in the following. The calculations were done using the so-called {\it graphical calculus}, a framework that is briefly outlined in Appendix \ref{G.Cal.} along with the chosen notations and conventions.

Consider again the spin network state $\ket{\gamma,\{j\},\{\iota\}}$ containing the $n$-valent vertex $v$ ($n>1$), from which originate the two edges $e$ and $e'$. The actions of the Euclidean and Lorentzian parts of $\widehat{H_v^T}_{e,e'}$ on the state $\ket{v;j_e,j_{e'},\dots ;k_{e'},\dots}$, defined in appendix \ref{G.Cal.}, are respectively
\begin{align}
&\widehat{H_v^E}_{e,e'} \ket{v;j_e,j_{e'},\dots ;k_{e'},\dots} = \widehat{H_v^E}_{e,e'} \quad \makeSymbol{\includegraphics[scale=1.25]{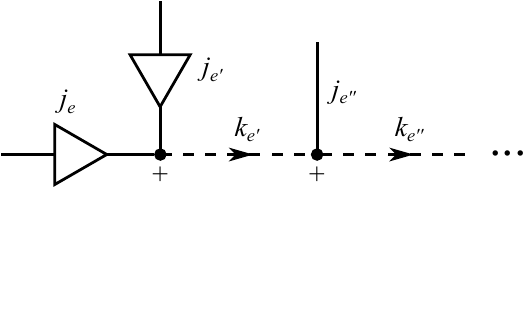}} \label{Eucl_ac} \\ 
 \nonumber &=\ \frac{3}{2}\sqrt{6}(-)^{j_e-j_{e'}-k_{e'}+1}\ \frac{i\ W_{j_e}W_{j_{e'}}}{W_l}\ \kappa_1(v)\\ \nonumber &\quad \times \sum_{x_e} d_{x_e} 
\begin{Bmatrix}
 j_{e'} & j_{e'} & 1 \\
 j_e & {x_e} & k_{e'}
\end{Bmatrix}
\begin{Bmatrix}
 1 & 1 & 1 \\
 j_e & {x_e} & j_e
\end{Bmatrix}
\quad \makeSymbol{\includegraphics[scale=1]{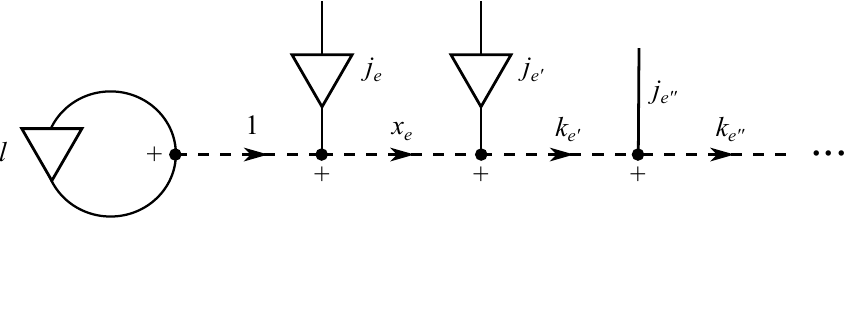}}\ ,
\end{align}

and
{\small
\begin{align}\label{Loren_ac}
\widehat{H_v^L}_{e,e'} \ket{v;j_e,j_{e'},\dots;k_{e'},\dots}\ =\ &\frac{\kappa_2(v)}{2} \sqrt{\delta_{ii'} (\epsilon_{ijk} \hat{J}_{v,e,j} \hat{J}_{v,e',k}) (\epsilon_{i' j' k'} \hat{J}_{v,e,j'} \hat{J}_{v,e',k'})} \\ \nonumber
\ &\left(\pi - \arccos \left[ \frac{ \delta_{kl} \hat{J}_{v,e,k} \hat{J}_{v,e',l}}{\sqrt{\delta_{kk'} \hat{J}_{v,e,k} \hat{J}_{v,e,k'}} \sqrt{\delta_{ll'} \hat{J}_{v,e',l} \hat{J}_{v,e',l'}}} \right]\right)\ket{v;j_e,j_{e'},\dots;k_{e'},\dots}\\ \nonumber \\ \nonumber
=\ &\frac{\kappa_2(v)}{2} \sqrt{c(j_e,j_{e'},k_{e'})} \left( \frac{2\pi}{\alpha_{v,e,e'}} - \theta(j_e,j_{e'},k_{e'}) \right) \ket{v;j_e,j_{e'},\dots;k_{e'},\dots},
\end{align}
}

where
{\small
\begin{align}
\nonumber c(j_e,j_{e'},k_{e'}):=\ & j_e(j_e+1)j_{e'}(j_{e'}+1)-\left(\frac{k_{e'}(k_{e'}+1)-j_e(j_e+1)-j_{e'}(j_{e'}+1)}{2}\right)^2\\ \quad & -\frac{k_{e'}(k_{e'}+1)-j_e(j_e+1)-j_{e'}(j_{e'}+1)}{2},
\end{align}
\begin{align}
\theta(j_e,j_{e'},k_{e'}):=\ \pi - \arccos \left[ \frac{ k_{e'} (k_{e'}+1) - j_e (j_e+1) - j_{e'} (j_{e'}+1)} {2 \sqrt{j_e (j_e+1) j_{e'} (j_{e'}+1)} } \right].
\end{align}
}

Notice that while the operator $\widehat{H_v^E}_{e,e'}$ in the Euclidean part is a graph changing operator, hence not preserving the original intertwiner space, the operator $\widehat{H_v^L}_{e,e'}$ in the Lorentzian part is diagonal on the basis adapted to the pair of edges $\{e,e'\}$. Also, from equations (\ref{Eucl_ac}) and (\ref{Loren_ac}), we can deduce that the domain of the Hamiltonian operator $\hat{H}$ admits an orthogonal sum decomposition in terms of stable subspaces under repeated action of $\hat{H}$. This result generalizes to other symmetrizations than the one proposed in (\ref{Sym.Hamil}), and it may be of considerable importance in the elaboration of self-adjointness proofs and the calculation of the evolution of physical states in this model.

\section{Summary and outlooks}\label{section_3}

We considered a model of Einstein gravity coupled to a free scalar field, in which the dynamics of the gravitational field is described by deparametrization with respect to the scalar field. In the corresponding quantum theory, constructed using the techniques of loop quantum gravity, the quantum dynamics is given by the evolution of the physical (i.e. gauge and diffeomorphism invariant) states of the gravitational field with respect to the scalar field. This evolution is governed by a physical Hamiltonian operator, which we constructed in this paper. The implementation of the Lorentzian part of our Hamiltonian is based on regularization used to define the curvature operator introduced in \cite{Curvature_op.}. As to the Euclidean part, we refined and made precise the idea, first considered in \cite{Rov-Smo}, of regularizing the curvature by means of loops attached to pairs of edges at a vertex of a spin network graph.

By carefully specifying the properties of the {\em special loops} created by the Euclidean Hamiltonian operator, we were able to define an operator which is diffeomorphism invariant, and whose adjoint operator is densely defined. The second property is crucial in that it allows to symmetrize the operator and eventually to construct self-adjoint extensions. Our regularization of the Euclidean term can also be applied in vacuum loop quantum gravity to define a Hamiltonian constraint operator for which the adjoint operator is densely defined. This question will be treated in a future work. On a practical level, an important feature of our Hamiltonian is that the volume operator does not appear in it. This implies a considerable simplification of the calculation of the action of the Hamiltonian on spin network states.

The construction presented in this paper gives us a concrete and tractable Hamiltonian operator for loop quantum gravity coupled to a free scalar field. This makes it possible to test the dynamics of the theory, as the time evolution of spin networks under this Hamiltonian can be computed. In particular, a question of interest will be to study the evolution of semiclassical states describing e.g. cosmological spacetimes.

\begin{center}
 \large{\bf{Acknowledgments}}
\end{center}
We would like to thank Yongge Ma for discussions and comments about the quantization of the physical Hamiltonian considered in this paper. We also benefited a lot from discussions with Hanno Sahlmann and Wojciech Kaminski. This work was supported by the grant of Polish Narodowe Centrum Nauki nr 2011/02/A/ST2/00300. I.M. would like to thank the Jenny and Antti Wihuri Foundation for support.\\

\appendix

\section{Graphical calculus for $SU(2)$}\label{G.Cal.}

In this appendix we give the technical tools of $SU(2)$ representation theory that are needed to evaluate the action of the Hamiltonian on a spin network state. For a more detailed presentation, we refer the reader to \cite{BrinkSatchler}.

\subsection{The epsilon tensor}

The epsilon tensor is the fundamental invariant tensor of $SU(2)$. For the spin-$j$ representation, it is given in the standard basis\footnote{All the relations and conventions given in this appendix refer to the basis $\{\ket{j,m}\}$, which is the usual eigenbasis of the operators $J^2$ and $J_3$. For $j=\tfrac{1}{2}$, the basis $\{\ket{\tfrac{1}{2},+{}\tfrac{1}{2}}, \ket{\tfrac{1}{2},-{}\tfrac{1}{2}}\} \equiv \{\ket +,\ket -\}$ of the space ${\cal H}_{1/2}$ is defined by specifying that $\tau_3\in su(2)$ and $\epsilon\in {\cal H}_{1/2}^*\otimes{\cal H}_{1/2}^*$ are represented by the following matrices:
$$
(\tau_3)^A_{\ B} = -\frac{i}{2}\begin{pmatrix} 1&0 \\ 0&-1 \end{pmatrix}, \qquad \epsilon_{AB} = \begin{pmatrix} 0&1 \\ -1&0 \end{pmatrix},
$$
the indices $A$ and $B$ taking the values $+$ and $-$. For a general representation $j$, the normalized states $\ket{j,m}$, which span the space ${\cal H}_j$, are obtained as symmetric tensor products of the states $\ket +$ and $\ket -$:
$$
\ket{j,m} = \sqrt{\frac{(j+m)!(j-m)!}{(2j)!}}\Bigl(\underbrace{\ket +\otimes\cdots\otimes\ket +}_{\text{$j+m$ states}} \otimes\underbrace{\ket -\otimes\cdots\otimes\ket -}_{\text{$j-m$ states}} \; +\; \text{all permutations}\Bigr).
$$
}  $\{\ket{j,m}\}$ by
\begin{equation}
\epsilon^{(j)}_{mn} = (-1)^{j-m}\delta_{m,-n}
\end{equation}
and is represented graphically as
\be
\epsilon^{(j)}_{mn} \ = \ \makeSymbol{\includegraphics[scale=1.25]{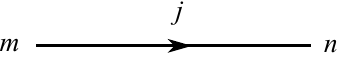}}
\ee
It satisfies the symmetry relation $\epsilon^{(j)}_{nm} = (-1)^{2j}\epsilon^{(j)}_{mn}$, i.e.
\begin{equation}
\makeSymbol{\includegraphics[scale=1.25]{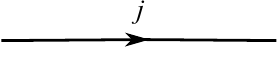}} \ = \ (-1)^{2j}\makeSymbol{\includegraphics[scale=1.25]{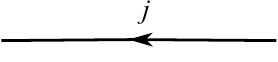}}
\end{equation}
The tensor $\epsilon^{(j)mn}$ is defined to be numerically equal to $\epsilon^{(j)}_{mn}$. The contraction of two epsilons gives
\begin{equation}
\epsilon^{(j)}_{m\mu}\epsilon^{(j)n\mu} = \delta_m^n.
\end{equation}
Graphically, $\delta_m^n$ is represented by a line with no arrow, and the above relation reads
\begin{equation}
\makeSymbol{\includegraphics[scale=1.25]{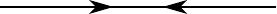}} \ = \ \makeSymbol{\includegraphics[scale=1.25]{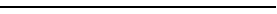}}
\end{equation}
Indices of $SU(2)$ tensors can be raised and lowered using the epsilon tensor. Our convention is the following:
\begin{equation}
v^m = \epsilon^{mn}v_n, \qquad v_m = v^n\epsilon_{nm}.
\end{equation}

\subsection{Intertwiners}

The intertwiner between three representations $j_1$, $j_2$ and $j_3$ is given by the Wigner 3$j$-symbol:
\begin{equation}\label{iota3}
\iota_{m_1m_2m_3} = \begin{pmatrix} j_1&j_2&j_3 \\ m_1&m_2&m_3 \end{pmatrix} = \makeSymbol{\includegraphics[scale=1.25]{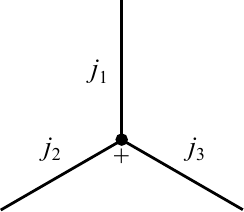}}
\end{equation}
The order of the spins in the symbol is indicated by a $+$ or a $-$ at the node. Thus,
\begin{equation}
\makeSymbol{\includegraphics[scale=1.25]{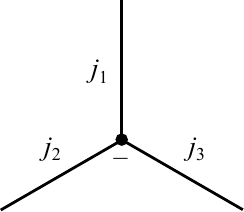}} \ = \ (-1)^{j_1+j_2+j_3}\makeSymbol{\includegraphics[scale=1.25]{3j.eps}}
\end{equation}
which is a graphical representation of the relation that switching two columns in the symbol \eqref{iota3} multiplies the symbol by $(-1)^{j_1+j_2+j_3}$. Another symmetry relation is
\begin{equation}
\makeSymbol{\includegraphics[scale=1.25]{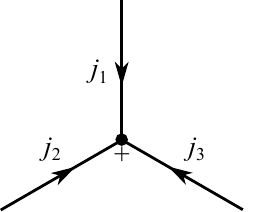}} = \makeSymbol{\includegraphics[scale=1.25]{3j.eps}}
\end{equation}
When one of the spins is zero, the 3$j$-symbol reduces to the epsilon tensor:
\begin{equation}
\makeSymbol{\includegraphics[scale=1.25]{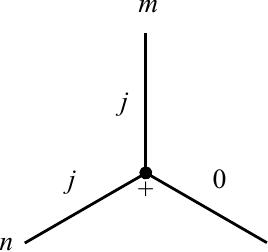}} \ = \ \frac{1}{\sqrt{d_j}}\;\makeSymbol{\includegraphics[scale=1.25]{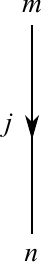}}
\end{equation}
where $d_j= 2j+1$ is the dimension of the representation $j$. Intertwiners of higher valence are constructed by contracting several three-valent intertwiners. For example, the objects
\begin{equation}\label{iota4}
\iota^{(k)}_{m_1m_2;m_3m_4} = \sum_\mu \begin{pmatrix} j_1&j_2&k \\ m_1&m_2&\mu \end{pmatrix}(-1)^{k-\mu}\begin{pmatrix} k&j_3&j_4 \\ -\mu&m_3&m_4 \end{pmatrix}
= \makeSymbol{\includegraphics[scale=1.25]{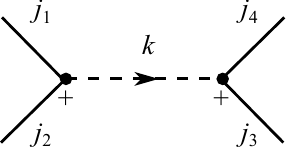}}
\end{equation} 
form a basis in the space of intertwiners between the representations $j_1$, $j_2$, $j_3$ and $j_4$. (Note that $\iota^{(k)}_{m_1m_2;m_3m_4}$ is not normalized; its norm is $1/\sqrt{d_k}$.)

In calculating the action of the Hamiltonian, a specific basis of intertwiner states at the vertex $v$ is chosen. These states are denoted by $\ket{v; j_e,j_{e'},j_{e''},\dots; k_{e'},k_{e''},\dots}$ and defined as
\be
\ket{v; j_e,j_{e'},j_{e''},\dots; k_{e'},k_{e''},\dots} \ := \ \makeSymbol{\includegraphics[scale=1.25]{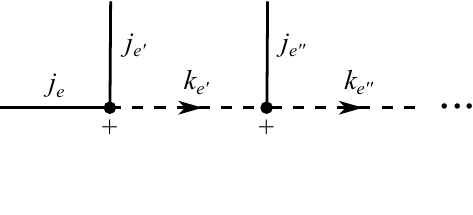}}.
\ee

\subsection{$6j$- and $9j$-symbols}

A contraction of four 3$j$-symbols defines the 6$j$-symbol,
\begin{align}
\begin{Bmatrix} j_1&j_2&j_3 \\ k_1&k_2&k_3 \end{Bmatrix}
\ = \makeSymbol{\includegraphics[scale=1.25]{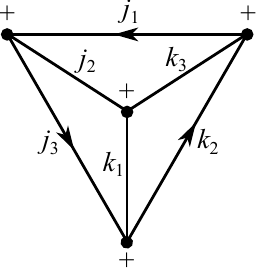}}
\end{align}
This object appears in the following relation, which is a special case of equation \eqref{expansion4} below, and which gives the change of basis between two different bases of the form \eqref{iota4}:
\begin{equation}
\makeSymbol{\includegraphics[scale=1.25]{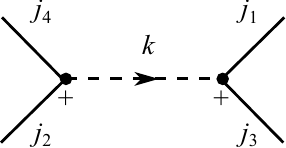}} = \sum_x d_x(-1)^{j_1+j_4-k-x}\begin{Bmatrix} j_1&j_2&x \\ j_4&j_3&k\end{Bmatrix} \makeSymbol{\includegraphics[scale=1.25]{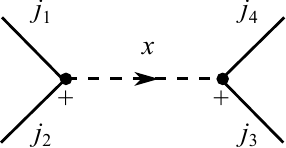}}
\end{equation}
Similarly, the 9$j$-symbol is defined as
\begin{equation}
\begin{Bmatrix} j_1&j_2&j_3 \\ k_1&k_2&k_3 \\ l_1&l_2&l_3 \end{Bmatrix} \ = \ \makeSymbol{\includegraphics[scale=1.25]{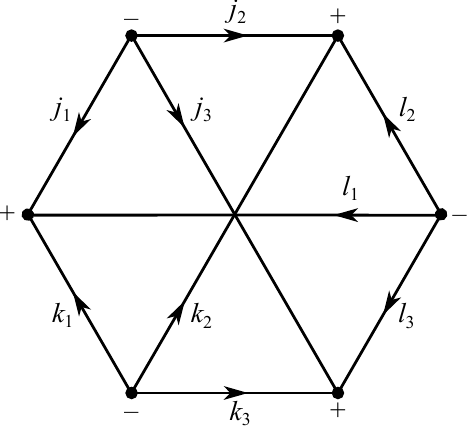}}
\end{equation}
It can be expressed in terms of 6$j$-symbols by the relation
\begin{equation}
\begin{Bmatrix} j_1&j_2&j_3 \\ k_1&k_2&k_3 \\ l_1&l_2&l_3 \end{Bmatrix} = \sum_x d_x(-1)^{2x} \begin{Bmatrix} j_1&j_2&j_3 \\ k_3&l_3&x \end{Bmatrix} \begin{Bmatrix} k_1&k_2&k_3 \\ j_2&x&l_2 \end{Bmatrix} \begin{Bmatrix} l_1&l_2&l_3 \\ x&j_1&k_1 \end{Bmatrix}.
\end{equation}

\subsection{Expanding invariant tensors}

Any invariant tensor $t_{m_1\cdots m_N}$, having indices in representations $j_1,\dots,j_N$, is an element of the space of intertwiners between the representations $j_1,\dots,j_N$, and as such, it can be expanded using a basis of the intertwiner space. Expressing a tensor with $N$ indices as a block to which $N$ lines are attached, one has the relations
\begin{align}
\makeSymbol{\includegraphics[scale=1.25]{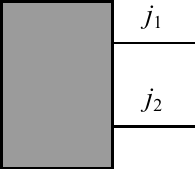}} \ &= \ \delta_{j_1,j_2}\frac{1}{d_{j_1}}\;\makeSymbol{\includegraphics[scale=1.25]{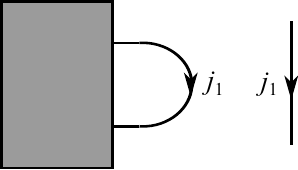}} \\
\makeSymbol{\includegraphics[scale=1.25]{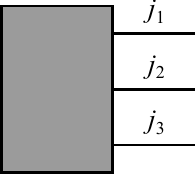}} \ &= \ \makeSymbol{\includegraphics[scale=1.25]{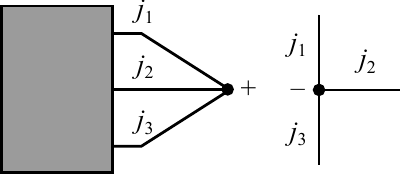}} \\
\makeSymbol{\includegraphics[scale=1.25]{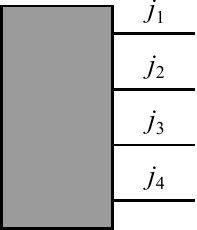}} \ &= \ \sum_x d_x\;\makeSymbol{\includegraphics[scale=1.25]{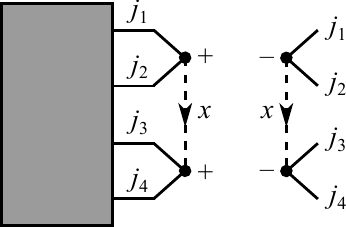}} \label{expansion4}
\end{align}
as well as the straightforward generalization of the last relation for tensors of higher order.

\subsection{Group elements}

The representation matrix for a group element is expressed graphically as
\begin{equation}
{D^{(j)}}^m_{\ \ n}(g) \ = \ \makeSymbol{\includegraphics[scale=1.25]{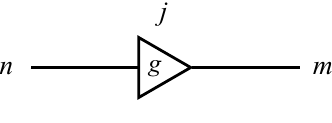}}
\end{equation}
The inverse matrix is given by the relation
\begin{equation}
{D^{(j)}}^m_{\ \ n}(g^{-1}) = \epsilon^{(j)m\mu}\epsilon^{(j)}_{n\nu}{D^{(j)}}^\nu_{\ \mu}(g),
\end{equation}
or
\begin{equation}
\makeSymbol{\includegraphics[scale=1.25]{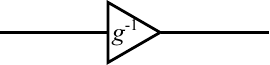}} \ = \ \makeSymbol{\includegraphics[scale=1.25]{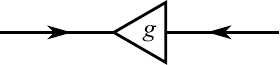}}
\end{equation}

In computing the action of the Hamiltonian, we need to know the action of a flux operator $\hat P_{S,i}$ on a holonomy $h_e$. In the cases where the intersection $v$ between the surface $S$ and the edge $e$ is the beginning or ending point of the edge, this action is given by
\begin{align}\label{actionofflux}
\hat P_{S,i}{D^{(j)}}^m_{\ \ n}(h_e) &= \frac{1}{2} \kappa(S,e) \hat J_{v,e,i} {D^{(j)}}^m_{\ \ n}(h_e)\\ \nonumber &= \frac{i}{2}\kappa(S,e)\times\begin{cases} {D^{(j)}}^m_{\ \ \mu}(h_e)(\tau_i^{(j)})^\mu_{\ \ n} & \text{if $e$ begins from $v$} \\ (\tau_i^{(j)})^m_{\ \ \mu}{D^{(j)}}^\mu_{\ n}(h_e) & \text{if $e$ ends on $v$} \end{cases}
\end{align}
where $\kappa(S,e) = +1$ if the orientation of $e$ agrees with the direction of the normal vector of $S$, and $\kappa(S,e) = -1$ if the orientations of $e$ and $S$ are opposite.

The matrix $\tau_i^{(j)}$ is proportional to an intertwiner between the representations $j$, $j$ and 1; the precise relation is
\vspace{-36pt}
\begin{equation}
(\tau_i^{(j)})^m_{\ \ n} \ = \ W_j\;\makeSymbol{\includegraphics[scale=1.25]{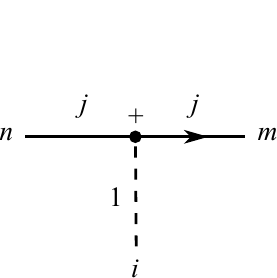}}
\end{equation}
where $W_j = i\sqrt{j(j+1)(2j+1)}$. Therefore we can write \eqref{actionofflux} graphically as
\vspace{-30pt}
\begin{align}
\hat P_{S,i}\;\makeSymbol{\includegraphics[scale=1.25]{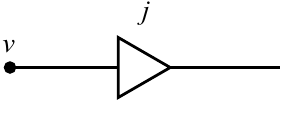}} \ &= \ \frac{i}{2}W_j\kappa(S,e)\;\makeSymbol{\includegraphics[scale=1.25]{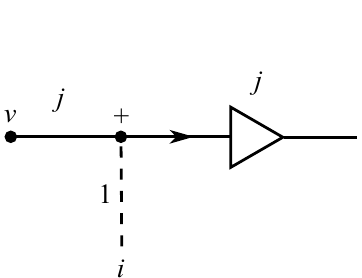}} \\
\hat P_{S,i}\;\makeSymbol{\includegraphics[scale=1.25]{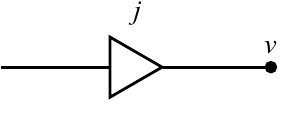}} \ &= \ \frac{i}{2}W_j\kappa(S,e)\;\makeSymbol{\includegraphics[scale=1.25]{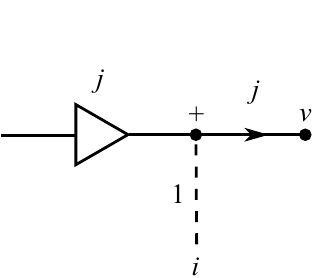}}
\end{align}

\bibliographystyle{plainnat}

\end{document}